\documentclass{aa}  

\usepackage{graphicx}
\usepackage{amsmath}	
\usepackage{booktabs}
\usepackage{newtxtext,newtxmath}
\usepackage{xcolor}
\usepackage{pdflscape}
\newcommand{\cii}{[C\,{\footnotesize II}]}
\newcommand{\hi}{H\,{\footnotesize I}}

\newcommand{\hb}{H$\beta$}
\newcommand{\lya}{Ly$\alpha$}
\newcommand{\oiii}{[O\,{\footnotesize III}]}

\begin{document} 

   \title{Unveiling \cii\, clumps in a lensed star-forming galaxy at $z \sim 3.4$}
   
   \author{A. Zanella\inst{1}
          \and
          E. Iani\inst{2}
          \and
          M. Dessauges-Zavadsky\inst{3}
          \and
          J. Richard\inst{4}
          \and
          C. De Breuck\inst{5}
          \and
          J. Vernet\inst{5}
          \and
          M. Kohandel\inst{6}
          \and
          F. Arrigoni Battaia\inst{7}
          \and
          A. Bolamperti\inst{8,1,5}
          \and
          F. Calura\inst{9}
          \and
          C.-C. Chen\inst{10}
          \and
          T. Devereaux\inst{8}
          \and
          A. Ferrara\inst{6}
          \and
          V. Mainieri\inst{5}
          \and
          A. Pallottini\inst{6}
          \and
          G. Rodighiero\inst{8}
          \and
          L. Vallini\inst{9}
          \and
          E. Vanzella\inst{9}
          }

   \institute{Istituto Nazionale di Astrofisica (INAF), Vicolo dell'Osservatorio 5, I-35122 Padova, Italy\\
              \email{anita.zanella@inaf.it}
    \and
    Kapteyn Astronomical Institute, University of Groningen, NL-9700 AV Groningen, the Netherlands
    \and
    D\'epartement d'Astronomie, Universit\'e de Gen\`eve, Chemin Pegasi 51, CH-1290 Versoix, Switzerland
    \and
    Univ. Lyon, Univ. Lyon1, ENS de Lyon, CNRS, Centre de Recherche Astrophysique de Lyon UMR5574, 69230, Saint-Genis-Laval, France
    \and
    European Southern Observatory, Karl Schwarzschild Strasse 2, 85748, Garching, Germany
    \and
    Scuola Normale Superiore, Piazza dei Cavalieri 7, 56126 Pisa, Italy
    \and
    Max-Planck-Institut fur Astrophysik, Karl-Schwarzschild-Str 1, 85748, Garching bei Munchen, Germany
    \and
    Dipartimento di Fisica e Astronomia, Università degli Studi di Padova, Vicolo dell'Osservatorio 3, I-35122 Padova, Italy
    \and
    INAF - Osservatorio Astronomico di Bologna, Via Gobetti 93/3, I-40129 Bologna, Italy
    \and
    Academia Sinica Institute of Astronomy and Astrophysics (ASIAA), No. 1, Sec. 4, Roosevelt Road, Taipei 10617, Taiwan}

   \date{Received XXX; accepted XXX}


 
 \abstract
   {Observations at UV and optical wavelengths have revealed that galaxies at $z \sim 1 - 4$ host star-forming regions, dubbed ``clumps,'' which are believed to form due to the fragmentation of gravitationally unstable, gas-rich disks. However, the detection of the parent molecular clouds that give birth to such clumps is still possible only in a minority of galaxies, mostly at $z \sim 1$.}
   {We investigated the \cii\, and dust morphology of a $z \sim 3.4$ lensed galaxy hosting four clumps detected in the UV continuum. We aimed to observe the \cii\, emission of individual clumps that, unlike the UV, is not affected by dust extinction, to probe their nature and cold gas content.}
   {We conducted ALMA observations probing scales down to $\sim 300$ pc and detected three \cii\, clumps. One (dubbed ``NE'') coincides with the brightest UV clump, while the other two (``SW'' and ``C'') are not detected in the UV continuum. We do not detect the dust continuum.}
   {We converted the \cii\, luminosity of individual clumps into molecular gas mass and found $\mathrm{M_{mol} \sim 10^8\, M_\odot}$. By complementing it with the star formation rate (SFR) estimate from the UV continuum, we estimated the gas depletion time ($t_\mathrm{dep}$) of clumps and investigated their location in the Schmidt-Kennicutt plane. While the NE clump has a very short $t_\mathrm{dep} = 0.16\, \mathrm{Gyr}$, which is comparable with high-redshift starbursts, the SW and C clumps instead have longer $t_\mathrm{dep} > 0.65\, \mathrm{Gyr}$ and are likely probing the initial phases of star formation. The lack of dust continuum detection is consistent with the blue UV continuum slope estimated for this galaxy ($\beta \sim -2.5$) and it indicates that dust inhomogeneities do not significantly affect the detection of UV clumps in this target.}
   {We pushed the observation of the cold gas content of individual clumps up to $z \sim 3.4$ and showed that the \cii\, line emission is a promising tracer of molecular clouds at high redshift, allowing the detection of clumps with a large range of depletion times.}


   \maketitle
%

\section{Introduction}

In the last decades, rest-frame ultraviolet (UV) and optical observations have shown that star-forming galaxies at redshift $z \sim 1 - 4$ have irregular morphologies (e.g.,\citealt{Conselice2004, Conselice2014, Shibuya2016, Huertas-Company2023}), dominated by active sites of star formation, dubbed ``clumps'' (e.g., \citealt{Elmegreen2005, Elmegreen2008, ForsterSchreiber2011, Guo2015, Zanella2015}). Spatially resolved observations taken with ground-based adaptive optics facilities (e.g., SINFONI on the Very Large Telescope, VLT), the \textit{Hubble} Space Telescope (\textit{HST}), and more recently the \textit{James Webb} Space Telescope (\textit{JWST}), revealed that clumps typically have stellar masses $\mathrm{M_\star \sim 10^7 - 10^9\, M_\odot}$, star formation rates $\mathrm{(SFRs) \sim 0.1 - 10\, M_\odot yr^{-1}}$, and mostly unresolved sizes $< 1$ kpc (e.g., \citealt{ForsterSchreiber2011, Guo2018, Zanella2019, Kalita2023}). By combining the angular resolution of state-of-the-art telescopes with strong lensing, it has been possible to study clumps in the low-mass and low-SFR regime \citep{Livermore2015, Vanzella2017b, Vanzella2017a, Cava2018, Vanzella2021}. Such studies have revealed that when magnification (and hence spatial resolution and sensitivity) increases, clumps with smaller sizes are uncovered \citep{Vanzella2022, Mestric2022, Claeyssens2023, Messa2022}. In particular, clumps in lensed galaxies have effective radii $\mathrm{R_e \sim 10 - 100\, pc}$, stellar masses $\mathrm{M_\star \sim 10^6 - 10^8\, M_\odot}$, and $\mathrm{SFRs \sim 0.01 - 10\, M_\odot yr^{-1}}$ \citep{Mestric2022, Claeyssens2023}. They have blue UV continuum $\beta$ slopes, in several cases approaching extreme values ($\beta \sim -3$), indicating that they are active sites of star formation hosting young stellar populations with relatively low metallicity \citep{Bolamperti_prep}. Observational evidence suggests that the majority of $z \sim 1 - 4$ clumps have likely originated in situ in galaxies' disk \citep{Shibuya2016, Zanella2019}, supporting simulation results showing that clumps form due to disk instabilities in gas-rich galaxies \citep{Bournaud2014, Ceverino2015, Tamburello2015, Mandelker2017, Leung2020, Zanella2021}. If this scenario is indeed correct, we should also detect the parent molecular clouds that give birth to the clumps. 

The molecular hydrogen (H$_2$), the fuel for star formation, is not directly observable at high redshift and carbon monoxide (CO) is typically used as a molecular gas tracer instead. CO in local star-forming galaxies is organized in giant molecular clouds (GMCs), with typical sizes $\sim 5 - 100$ pc and masses $\mathrm{M_{H2} \sim 10^4 - 10^7\, M_\odot}$ \citep{Bolatto2013, Miville-Deschenes2017, Freeman2017}. Spatially resolved observations of the CO emission in $z \sim 1 - 3$ star-forming galaxies instead are still sparse. Simulations predict the CO(5-4) transition to be the brightest in clumps at $z \sim 1 - 4$ due to their reservoir of warm, dense gas \citep{Bournaud2015}. Observations have been performed in a clumpy galaxy at $z \sim 1.5$ \citep{Cibinel2015} and CO(5-4) has been detected from the nuclear region, coincident with a red, proto-bulge component. No detection has been found at the location of the star-forming clumps instead. This might be due to physical reasons, such as the short gas depletion timescale of clumps \citep{Cibinel2015}. In addition, observational limitations might be at play, such as the relatively coarse resolution of the observations ($\sim 1$ kpc) and the lack of sufficient sensitivity. In addition, to overcome the lack of spatial resolution and sensitivity, mostly bright submillimeter galaxies (SMGs) lensed by foreground galaxy clusters have been targeted with the Atacama Large Millimeter/submillimiter Array (ALMA) to detect the dust-continuum, CO, or \cii\, line emission of individual high-redshift clumps. Most of these works led to tentative or non-detections, or morphologies consistent with homogeneous dust and gas distributions \citep{Gullberg2015, Hodge2016, Canameras2017, Gullberg2018, Tadaki2018, Rujopakarn2019, Hodge2019, Ivison2020, Ushio2021, Calura2021, Liu2023}. However, since rest-frame UV observations of such SMGs are not available, it is unclear whether these targets are clumpy in the first place. The first robust detections of the CO(4-3) emission from individual clouds have been obtained in two clumpy, main-sequence galaxies at $z \sim 1$, lensed by foreground galaxy clusters, the Cosmic Snake \citep{Dessauges-Zavadsky2019} and A521 \citep{Dessauges-Zavadsky2023}. Thanks to the combination of high-resolution observations ($\lesssim 0.2\arcsec$) and lensing magnification, tens of GMCs with sizes $\sim 30 - 200$ pc have been identified. They have molecular gas masses 100 times higher than local GMCs and ten times higher molecular gas surface densities ($\mathrm{\Sigma_{gas} \sim 10^3 - 10^4\, M_\odot\, pc^{-2}}$, \citealt{Dessauges-Zavadsky2023}). The GMCs in the Cosmic Snake and A521 show a spatial offset with respect to the clumps detected at rest-frame UV wavelengths, possibly indicating that GMCs are quickly disrupted (tens of millions of years) or dispersed after the first episode of star formation \citep{Dessauges-Zavadsky2023}, similarly to local GMCs that have typical lifetimes of $\sim 10 - 30$ Myr \citep{Kruijssen2019, Chevance2020, Kim2021}.

To detect the gas reservoir of young clumps, we targeted a $z \sim 3.4$ galaxy lensed by the foreground cluster Abell 2895, hosting four UV-bright clumps. High-resolution observations were carried out with ALMA targeting the \cii\,$\lambda 158\mu m$ emission line. This far-infrared (far-IR) fine-structure line is one of the main coolants of the interstellar medium \citep[ISM, ][]{Stacey1991, Gracia-Carpio2011, Carilli2013} and it has been often considered as a SFR tracer \citep{deLooze2010, deLooze2014, Capak2015}. However, in  recent years, there has been increasing observational evidence that \cii\, is tightly correlated with the molecular gas \citep{Zanella2018, Madden2020, Dessauges-Zavadsky2020, Gururajan2023}, also supported by theoretical works and simulations \citep{Pallottini2017a, Ferrara2019, Sommovigo2021, Vizgan2022}. In this work we discuss the morphological analysis of the cold ISM of our lensed, clumpy target, determine the gas reservoir of individual clumps traced by \cii\, and, thanks to complementary UV continuum observations tracing clumps' SFR, we estimate their gas depletion time.

This paper is organized as follows: in Section \ref{sec:data}, we describe our target galaxy and the data set; in Section \ref{sec:analysis}, we discuss how we obtained the \cii\, and dust continuum maps and spectra, we describe the morphology of the galaxy as determined by considering different tracers and the detection of individual clumps; in Section \ref{sec:results}, we discuss the physical properties (molecular gas mass, star formation rate, star formation efficiency) that we derived for our target; in Section \ref{sec:discussion}, we report scaling relations among observables and place them in the context of current literature findings; finally, in Section \ref{sec:conclusions}, we conclude and summarize our findings. Throughout the paper we adopt a flat $\Lambda$CDM cosmology with $\Omega_\mathrm{M} = 0.3$, $\Omega_\mathrm{\Lambda} = 0.7$, and $\mathrm{H_0} = 70$ km s$^{-1}$ Mpc$^{-1}$. All magnitudes are AB magnitudes \citep{Oke1974} and we adopt a \citep{Chabrier2003} initial mass function, unless differently stated.

\section{Data}
\label{sec:data}


\subsection{Our target galaxy and ancillary data}
\label{subsec:target}

\begin{figure*}
    \centering
    \includegraphics[width=0.7\textwidth]{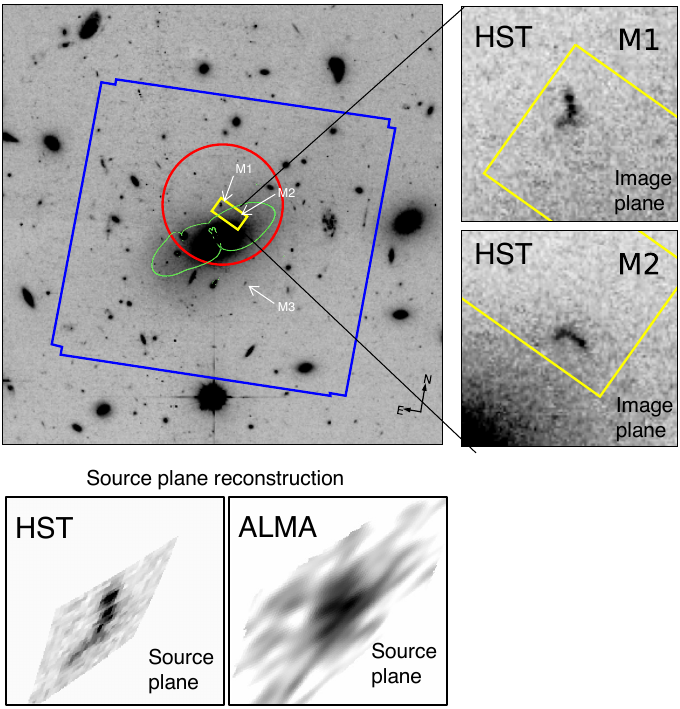}
    \caption{Available observations for our target galaxy. \textit{Left top panel}: the \textit{HST}/ACS WFC F606W observations are shown as the background image, whereas the contours show the MUSE WFM+AO (blue) and SINFONI (yellow) field of view. The ALMA primary beam is also shown (red circle). The SINFONI and ALMA observations cover only two of the three multiple images, namely M1 and M2. The green contours show the critical lines. \textit{Right panels:} M1 (top) and M2 (bottom) multiple images of our target galaxy as detected in \text{HST} data \citep{Livermore2015, Iani2021}. Both cutouts have a size of about $5\arcsec \times 5\arcsec$. \textit{Bottom panels:} Source plane reconstruction of our target. We show both the UV continuum from \textit{HST} data (left) and the \cii\, emission from ALMA data (right). Both cutouts have a size of $1\arcsec \times 1\arcsec$.}
    \label{fig:abell2895}
\end{figure*}

Our target is a clumpy galaxy lensed by the foreground brightest central galaxy (BCG) of the cluster Abell 2895. Three multiple images (M1, M2, and M3) have been identified and analyzed in previous studies \citep{Livermore2015, Iani2021}. However, only two of them are included in the ALMA primary beam (Table \ref{tab:log} and Section \ref{subsec:ALMA_data}), hence in the following we only focus on M1 (RA = 01:18:11.184, DEC = -26:58:03.826) and M2 (RA = 01:18:10.851, DEC = -26:58:07.608; Figure \ref{fig:abell2895}). The average lensing magnification of M1 and M2 is $\mu = 5.5 \pm 0.7$ and $\mu = 4.5 \pm 0.3$, respectively (Section \ref{subsec:lensing}, \citealt{Livermore2015, Iani2021}).

Our target has a redshift $z_\mathrm{opt} = 3.39535 \pm 0.00025$, as estimated from optical emission lines \citep{Iani2021} and the morphology of both its UV continuum and optical line (\hb, \oiii) emission is clumpy. At least four star-forming clumps are detected on top of the diffuse emission, likely coming from the underlying disk (Figure \ref{fig:abell2895}). Individual clumps have stellar masses $\mathrm{M_\star \lesssim 2\times 10^8\, M_\odot}$ and effective radii, as measured from their UV continuum emission, $\mathrm{R_e \lesssim 250\, pc}$ (magnification corrected, \citealt{Iani2021}).

We complement our ALMA observations with a suite of ancillary data and a lensing model (Section \ref{subsec:lensing}). The rest-frame UV \textit{HST} imaging has $\mathrm{FWHM} \sim 0.13\arcsec$ resolution, while the VLT/SINFONI spectra targeting \hb\, and \oiii\, are seeing limited ($\mathrm{FWHM} \sim 0.6\arcsec$). The VLT/MUSE observations taken with the Adaptive Optics (AO) Wide Field Mode (WFM) and targeting the \lya\, emission have $\mathrm{FWHM} \sim 0.4\arcsec$ resolution. The peak of the \lya\, emission is spatially offset with respect to the UV continuum and the optical emission lines \citep{Iani2021}. Such offset between the optical emission, which is co-spatial with the clumps, and the \lya\, emission seems to indicate the presence of dust and/or neutral gas that absorb and scatter the \lya. This offset is not due to astrometry issues, as the UV continuum from MUSE and \textit{HST} data are co-spatial, and the \textit{HST} astrometry is calibrated against Gaia DR3 \citep{Iani2021}.

\begin{figure*}[t!]
    \centering
    \includegraphics[width=\textwidth]{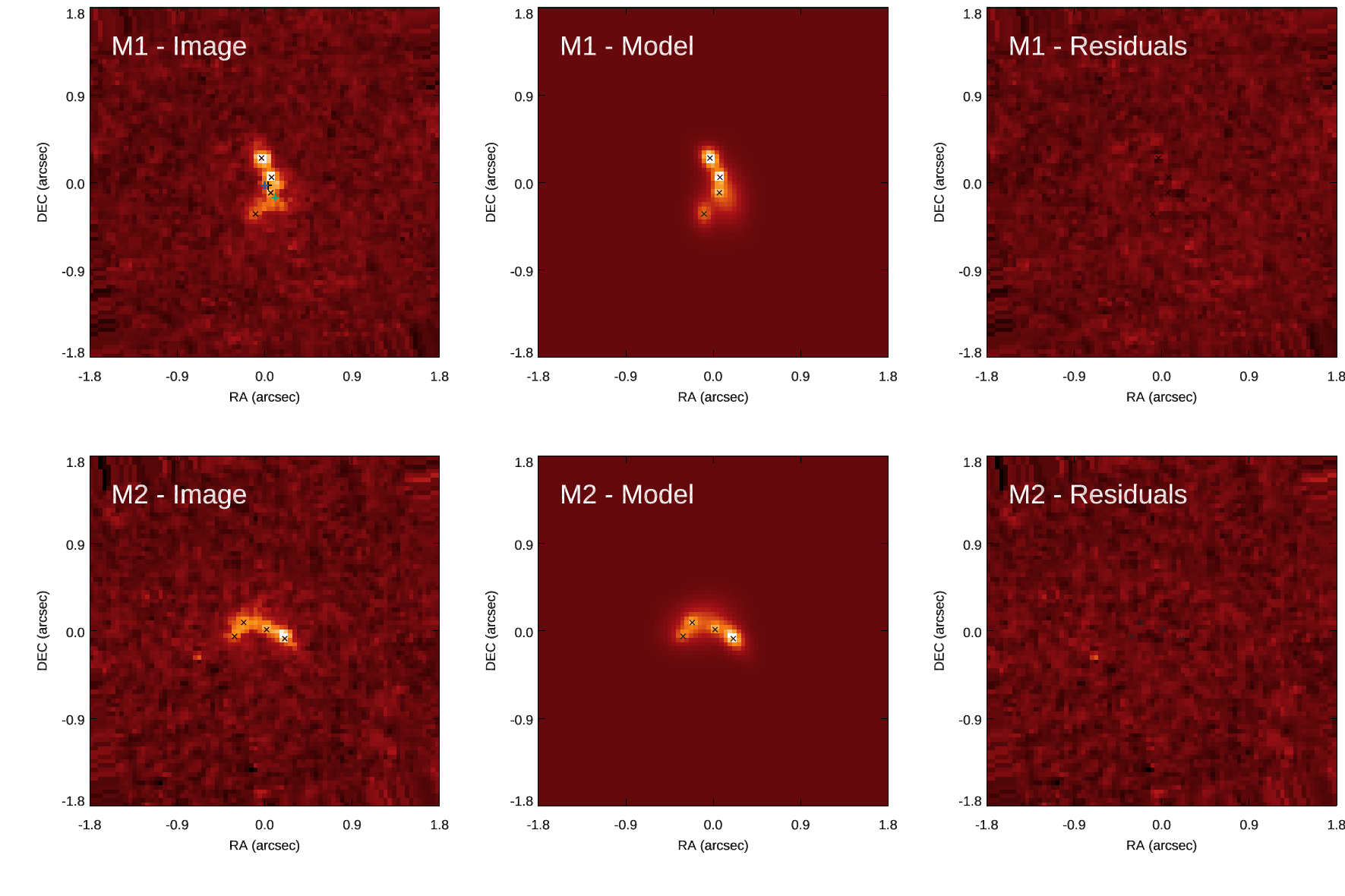}
    \caption{\textsc{Galfit} best-fit model of the UV continuum. We show the input image first column), the model (second column), and the residuals (third column) obtained by subtracting the model from the input image. Both galaxy images M1 (top row) and M2 (bottom row) are shown. The cutouts have a size of $3.6\arcsec \times 3.6\arcsec$. The color cuts are the same in all panels of M1 and in all panels of M2. The crosses in the top left panel indicate the barycenter of the M1 galaxy image as estimated with three different methods: from the \textsc{Galfit} disk-only best fit (black cross), from the \textsc{Galfit} disk+clumps best fit (cyan cross), and from the \textsc{Sextractor} (blue cross, see Section \ref{subsec:uv_model} for details). The x indicate the clumps detected in the UV continuum.}
    \label{fig:models}
\end{figure*}

\subsection{Lensing model}
\label{subsec:lensing}

The lensing model used in this work is discussed in detail in \cite{Iani2021}. In brief, we used a mass model constructed using the Lenstool\footnote{\url{https://projects.lam.fr/projects/lenstool/wiki}} software \citep{Jullo2007}, following the methodology described in \cite{Richard2010}. The large-scale and cluster structure 2D-projected mass distributions are modeled, respectively, as a parametric combination of a cluster-scale and multiple galaxy-scale double pseudo-isothermal elliptical potentials \citep{Eliasdottir2007}. The centers and shapes of the galaxy-scale components are constrained to the centroid, ellipticity and position angle of cluster members as measured on the \textit{HST} image, to limit the number of parameters in the model. The cluster members are selected through the color–magnitude diagram method (e.g. \citealt{Richard2014}) and we assume that they follow the Faber–Jackson relation for elliptical galaxies \citep{Faber1976}. This  model is constrained by using the location of our target and that of another triply imaged system with spectroscopic redshift $z \sim 3.7$ \citep{Livermore2015, Iani2023}. The best-fit model reproduces the location of the multiply imaged systems with a root-mean-square (rms) of 0.09\arcsec. With Lenstool we produce a 2D map of the magnification factor at the redshift of our target, we resample the magnification maps to match the spatial sampling of our data, and we reconstruct the multiple images on the source plane (Figure \ref{fig:abell2895}). This is done by using our lens model to raytrace back each spaxel, effectively subtracting the lensing displacement.

\begin{table}
    \centering
    \caption{Log of the ALMA observations.}
    \begin{tabular}{l c}
    \toprule
    \midrule
    ID                & A2895a \\
    Date              & 10 Oct 2019 \\
    t$_\mathrm{exp}$  & 1.9 hours \\
    Noise r.m.s. & 0.03 mJy/beam \\
    Beam          & $0.31\arcsec \times 0.26\arcsec$ \\  
    Observed frequency range & 418 - 434 GHz \\
    \bottomrule
    \end{tabular}
    \label{tab:log}
    
    \begin{minipage}{8.5cm}
    \textbf{Rows}: (1) Galaxy ID; (2) Date of observations; (3) Integration time on source; (4) Noise r.m.s. of the continuum, estimated over a bandwidth of 4811 km s$^{-1}$; (5) FWHM of the beam.
    \end{minipage}
\end{table}

\begin{table*}
    \centering
    \caption{Observed far-IR properties of our target galaxy and clumps.}
    \tiny
    \begin{tabular}{c c c c c c c c c}
    \toprule
    \midrule
    ID & RA & DEC & $z_\mathrm{\cii}$ & $\mu$F$_\mathrm{\cii}$ & $\Delta$v & $\mu$F$_\mathrm{cont}$ & $\mu$F$_\mathrm{UV}$ & $\mu$ \\
       &(deg) & (deg) &  & (mJy) & (km s$^{-1}$) & ($\mu$Jy) & ($\mathrm{10^{-31}erg\, s^{-1} cm^{-2} Hz^{-1}}$) & \\
    (1) & (2) & (3) & (4) & (5) & (6) & (7) & (8) & (9) \\
    \midrule
    Galaxy-integrated & 01:18:11.184 & -26:58:03.826 & $3.39548 \pm 0.00007$ & $7.5 \pm 0.7$ & 122 & $< 585.9$ & $63.3 \pm 3.2$ & $5.4 \pm 0.3$ \\
    Clump NE & 01:18:11.196 & -26:58:03.674 & - & $2.4 \pm 0.6$ & 61 & $< 190.0$ & $13.2 \pm 1.2$ & $5.1 \pm 0.1$ \\
    Clump SW & 01:18:11.169 & -26:58:03.952 & - & $2.5 \pm 0.6$ & 81 & $< 190.0$ & $< 2.1$ & $5.8 \pm 0.2$ \\
    Clump C (tentative) & 01:18:11.175 & -26:58:03.712 & - & $2.4 \pm 0.6$ & 41 & $< 190.0$ & $< 2.1$ & $5.3 \pm 0.1$ \\
    Clump UV-C1 & 01:18:11.1882 & -26:58:03.866 & - & - & - & - & $6.6 \pm 0.6$ & $5.4 \pm 0.1$ \\
    Clump UV-C2 & 01:18:11.1892 & -26:58:04.045 & - & - & - & - & $3.1 \pm 0.3$ & $5.7 \pm 0.1$ \\
    Clump UV-SE & 01:18:11.2028 & -26:58:04.290 & - & - & - & - & $10.6 \pm 1.0$ & $5.9 \pm 0.1$ \\
    \bottomrule
    \end{tabular}
    \label{tab:measurements}
    \vspace*{0.2cm}    
    \begin{minipage}{18cm}
    \textbf{Columns}: (1) Component: galaxy-integrated and individual clump measurements are reported. All the reported measurements refer to M1. While clumps NE, SW, and C are detected in \cii, clumps UV-C1, UV-C2, and UV-SE are only detected in the UV. (2) Right ascension; (3) Declination; (4) Redshift estimated by fitting the \cii\, emission line with a Gaussian in our 1D ALMA spectra. The uncertainty that we report is the formal error obtained from the fit; (5) Observed \cii\, emission line flux (i.e., not corrected for lensing amplification); (6) Velocity width of the line estimated considering the channels that maximize the S/N of the detection (see Section \ref{subsec:cii}); (7) 5$\sigma$ upper limits on the continuum emission flux (not corrected for lensing amplification). The upper limit associated to the galaxy-integrated measurement has been obtained by considering a Gaussian model with the same structural parameters used to estimate the integrated \cii\, flux (Section \ref{subsec:cii}). The upper limit associated to the clump measurements has instead been obtained by considering a point-spread function (PSF) model, for consistency with the \cii\, flux estimates; (8) observed UV continuum flux density. Upper limits are $5\sigma$; (9) average magnification \citep{Livermore2015, Iani2021}. \\
    \textbf{Note:} The coordinates of the \cii\, clumps are estimated as the centroid position of 2D fit performed in the \textit{uv} plane with GILDAS (Section \ref{subsec:cii}). We note that, when fitting relatively low S/N sources, GILDAS centroid positions might be affected by offsets, as reported by \cite{Tan2023}. However, the symmetry of the M1 and M2 images due to lensing makes the comparison of the relative position of the \cii\, and UV clumps robust (see Section \ref{subsubec:probab}).
    \end{minipage}
\end{table*}

\begin{table*}
    \centering
    \caption{Physical properties of our target galaxy and clumps. All properties have been corrected for magnification.}
    \begin{tabular}{c c c c c c c c}
    \toprule
    \midrule
    ID & $\mathrm{M_{mol}}$ & $\mathrm{SFR}$ & $\Sigma_\mathrm{mol}$ & $\Sigma_\mathrm{SFR}$ & $\mathrm{t_{dep}}$ & $\mathrm{R_{UV}}$ & $\mathrm{R_{\cii}}$ \\
       &($10^8\mathrm{M_\odot}$) & ($\mathrm{M_\odot yr^{-1}}$) & ($\mathrm{M_\odot pc^{-1}}$) & ($\mathrm{M_\odot yr^{-1} kpc^{-2}}$) & ($\mathrm{Gyr}$) & (pc) & (pc) \\
    (1) & (2) & (3) & (4) & (5) & (6) & (7) & (8) \\
    \midrule
    Galaxy-integrated & $20.5 \pm 2.0$ & $10.1 \pm 0.5$ & $734.5 \pm 59.9$ & $2.4 \pm 0.2$ & 0.38 $\pm$ 0.04 & 736 & $1352 \times 950$\\
    Clump NE & $3.5 \pm 0.9$ & $2.2 \pm 0.2$ & $1158.1 \pm 281.7$ & $7.3 \pm 0.1$ & 0.16 $\pm$ 0.04 & 251 & $< 311^\star$ \\
    Clump SW & $4.3 \pm 1.0$ & $< 0.3$ & $852.8 \pm 200.7$ & $< 0.6$ & $> 1.35$ & - & $< 399$ \\
    Clump C (tentative) & $2.2 \pm 0.5$ & $< 0.3$ & $408.5 \pm 100.0$ & $< 0.6$ & $> 0.65$ & - & $< 418$ \\
    Clump UV-C1 & - & $1.1 \pm 0.1$ & - & - & - & $< 207$ & - \\
    Clump UV-C2 & - & $0.5 \pm 0.04$ & - & - & $< 202$ & - \\
    Clump UV-SE & - & $1.6 \pm 0.1$ & - & - & $391$ & - \\
    \bottomrule
    \end{tabular}
    \label{tab:physical_prop}
    \vspace*{0.2cm}    
    \begin{minipage}{18cm}
    \textbf{Columns}: (1) Galaxy ID; (2) Molecular gas mass obtained from the \cii\, luminosity; (3) Star formation rate obtained from the UV luminosity; (4) Molecular gas surface density. For the \cii-detected clumps this is computed by considering the $\mathrm{R_{\cii}}$ upper limit as the size of the clump; (5) SFR surface density for the \cii-detected clumps. The $\mathrm{R_{\cii}}$ upper limit is considered as the size of the clump, for consistency with the $\Sigma_\mathrm{mol}$ estimate. If we were to consider the $\mathrm{R_{UV}}$ instead our conclusions would not change substantially; (6) Depletion time; (7) Effective radius of the UV continuum emission; (8) Semi-major and semi-minor axis of the \cii\, line emission, when resolved, and semi-minor axis upper limit (corresponding to the beam semi-minor axis) when unresolved. All properties listed here are intrinsic (i.e., corrected for magnification).\\
    $^\star$ This clump is detected also when using ``robust'' weighting with parameter 1 (instead of ``natural'' which is instead used to detect the other clumps). We determine the upper limit on the size of the NE clump from the semi-minor axis of the beam obtained with robust weighting, while for the SW and C clump we consider the beam size obtained with ``natural'' weighting. 
    \end{minipage}
\end{table*}

\begin{figure*}[t!]
    \centering
    \includegraphics[width=0.9\textwidth]{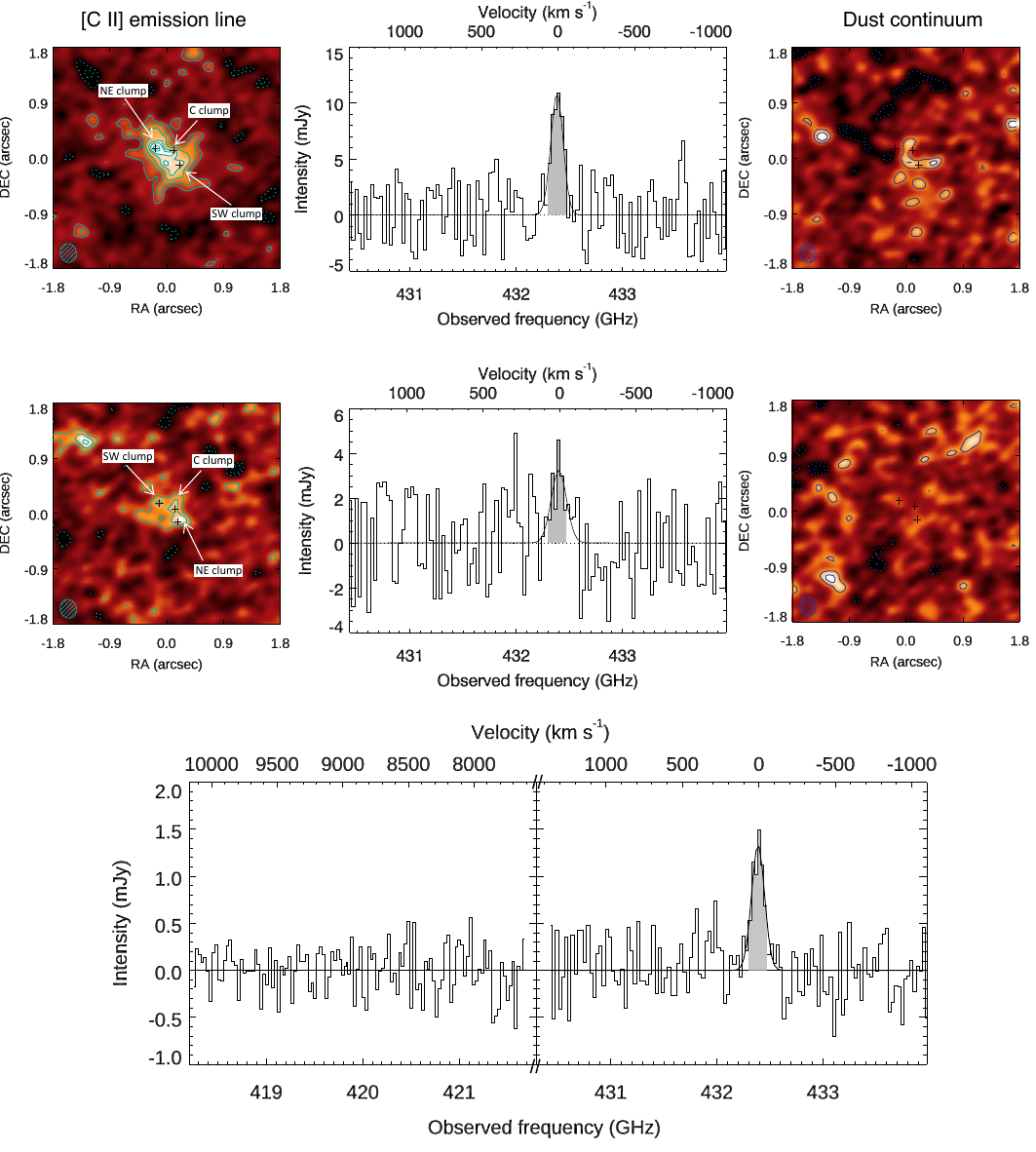}
    \caption{ALMA data of our galaxy. \textit{Top and middle rows}: we show both images covered by the primary beam, M1 (\textit{top row}) and M2 (\textit{middle row}). \textit{Left panels}: ALMA 2D maps of the \cii\, line. The cyan solid and dashed contours indicate respectively positive and negative levels of 2, 3, 4, 5, 5.5, 6 rms. The crosses indicate the location of the \cii\, clumps (NE, C, and SW as labeled on the figure). The beam is reported as the cyan ellipse. Each stamp has a size of $3.6\arcsec \times 3.6\arcsec$. \textit{Middle panels}: 1D spectrum from the ALMA datacube, extracted using the 2D Gaussian model that maximizes the S/N (Section \ref{subsec:cii}). The gray shaded areas indicate the velocity range over which we measured the \cii\, line flux. For illustrative purposes we also report the Gaussian fit of the emissions: it was not used to estimate the line fluxes, but only as an alternative estimate of the redshift of the galaxies (Section \ref{subsec:cii}). The intensity is observed, not corrected for magnification. \textit{Right panels}: ALMA 2D map of the continuum emission on the observed frequency range of 418 - 434\, GHz. The dark blue solid and dashed contours indicate respectively positive and negative levels from 2$\sigma$ to 4$\sigma$, in steps of 1$\sigma$. The intensity is observed, not corrected for magnification. \textit{Bottom row}: Intrinsic (magnification-corrected) stacked 1D spectrum of M1 and M2.}
    \label{fig:fig_maps}
\end{figure*}

\subsection{ALMA data}
\label{subsec:ALMA_data}

We carried out ALMA Band 8 observations for our target galaxy during Cycle 7 (PI: E. Iani, Project ID: 2019.1.01676.S) with the goal of detecting the \cii\, emission line, one of the brightest far-infrared cooling lines (rest-frame frequency $\mathrm{\nu_{rf} = 1905.1\ GHz}$), and possibly the underlying continuum. The dust continuum at these frequencies ($\mathrm{\nu_{obs} = 420 - 430\ GHz}$) provides direct information about the dust-obscured SFR.

We observed our target for 1.9 hours on source and reached a continuum sensitivity of 0.03 mJy/beam, over a bandwidth of 4811 km s$^{-1}$, and a sensitivity of 0.20 mJy/beam, over a bandwidth of 102 km s$^{-1}$, corresponding to the velocity width encompassed by the \cii\, emission (Section \ref{subsec:cii}). The native spectral resolution of the observations is 1.129 MHz ($\sim 0.78$ km s$^{-1}$), later binned to lower velocity resolution ($\sim 20.3$ km s$^{-1}$) to achieve sufficient S/N for our purposes. The imaged beam size (natural weighting) is $\mathrm{FWHM = 0.31\arcsec \times 0.26\arcsec}$ (Table \ref{tab:log}).
We reduced the data with the standard ALMA pipeline, based on the CASA software, version 5.6.1 \citep{McMullin2007}. The quasar J2258-2758 was chosen as the flux and bandpass calibrator. We then converted the calibrated datacubes to \textit{uvfits} format and analyzed them with the software GILDAS \citep{Guilloteau2000}. We did not perform continuum subtraction, as the continuum of our target is not detected (see Section \ref{subsec:continuum}).

\section{Analysis}
\label{sec:analysis}

\subsection{Modeling the UV continuum 2D light profile}
\label{subsec:uv_model}

 We fit the UV continuum emission of our target by using \textsc{GALFIT} \citep{Peng2002, Peng2010}. The process we adopted is fully detailed in \cite{Iani2021}, but we briefly summarize it in the following. We created a median PSF by stacking the non-saturated stars detected in the \textit{HST} field of view. We subtracted the BCG 2D light profile by fitting it with two S\'ersic models, convolved with the PSF. We then modeled the 2D surface brightness of our target with a 2D S\'ersic profile (the galaxy ``diffuse'' component) and subtracted the best-fit model from the data to obtain a ``residuals'' map. In the residuals we identified four clumps. We ran again \textsc{GALFIT} including additional parametric models to fit the clumps, namely two S\'ersic profiles for the clumps with larger diameter than the FWHM of the PSF and two PSF profiles for clumps that are spatially unresolved. Adopting Gaussian profiles instead of PSFs to model the clumps yields consistent results; however, we prefer to adopt PSF profiles to limit the number of free parameters of the fit. To estimate the intrinsic (magnification-corrected) flux of individual clumps, we divided the observed \textit{HST} image by the amplification map (Section \ref{subsec:lensing}) to obtain pixel-by-pixel corrected fluxes. We then fit the 2D light distribution of the galaxy with the best-fit model, letting the normalization (magnitude) of each component free to fit the data. Given that the magnification is fairly uniform for our target, we instead estimated the magnification-corrected effective radii of the marginally resolved clumps by dividing their observed effective radius (as derived by GALFIT) by the square-root of the average magnification factor at the location of the clump. The best-fit parameters are reported in Table \ref{tab:measurements}. 
 We repeated this analysis independently on M2 and reached consistent conclusions.
 
 In Figure \ref{fig:models}, we show the best-fit model and the residuals obtained by subtracting the model from the data for M1 and M2. The residuals are consistent with noise fluctuations and the center of the S\'ersic profile associated with the galaxy ``diffuse'' component is consistent ($< 0.2\arcsec$ difference) with the barycenter estimated with SExtractor \citep{Bertin1996}. To check the robustness of our estimate of the UV barycenter, we also smoothed the \textit{HST} data to the angular resolution of the ALMA data, by convolving them with a Gaussian kernel. The clumps are less prominent, but the UV emission still appears less spatially extended than the \cii. We estimated the barycenter of the UV emission by fitting the smoothed data with a single S\'ersic profile. Once again, the barycenter coordinates are consistent with those obtained with SExtractor (Figure \ref{fig:models}). In the following, we refer to the UV barycenter as the one derived with SExtractor.

\subsection{Continuum emission map}
\label{subsec:continuum}
We created averaged continuum maps by integrating the spectral range, after excluding the channels where the flux is dominated by the \cii\, emission line (Section \ref{subsec:cii}). We do not detect the continuum (see Figure \ref{fig:fig_maps}), which is consistent with the blue $\beta = -2.53 \pm 0.15$ slope of the galaxy as estimated from the rest-frame UV spectrum \citep{Iani2021}. In Table \ref{tab:measurements}, we report the total flux density upper limit that we estimated as the 5$\sigma$ uncertainty obtained when fitting, in the \textit{uv} plane, the Fourier Transform of a 2D Gaussian model with center and FWHM fixed at the position of the \cii\, detection. We also report the 5$\sigma$ upper limit obtained when fitting it with the Fourier Transform of a PSF model at the location of the clumps (see Section \ref{subsec:cii}).

\subsection{[C II] emission line map}
\label{subsec:cii}

To find the \cii\, emission line, determine the optimal channel range encompassing the emission, and create the velocity-integrated \cii\, emission map of our galaxy, we ran an iterative procedure, as the one described in several literature works (e.g., \citealt{Daddi2015, Zanella2018, Coogan2018, Zanella2023}). We briefly summarize it in the following.
We modeled our target's emission in the \textit{uv} plane with the GILDAS task \texttt{uv\_fit}, adopting a two-dimensional (2D) Gaussian profile in all four sidebands and channel per channel. To begin with, we fixed the spatial position to that determined from the optical \textit{HST} images (Section \ref{subsec:target}). Using the best-fit 2D Gaussian model we extracted the one-dimensional (1D) spectrum and we searched for a positive emission line signal in the resulting spectrum. We averaged the data over the channels, maximizing the detection signal-to-noise ratio (S/N) and we fit the resulting 2D (channel-averaged) map to obtain a new best-fitting line spatial position. If this was different from the spatial position of the previous extraction, we proceeded to a new spectral extraction at the new position, and iterated the procedure until convergence was reached.

We securely detected the \cii\, emission line of the most magnified image of our target, M1, at $\sim 10 \sigma$ significance (Figure \ref{fig:fig_maps}, Table \ref{tab:measurements}). 
We estimated the redshift from the \cii\, line in two ways, both giving consistent results (obtaining a discrepancy $\Delta z < 0.00013$): by computing the signal-weighted average frequency within the line channels and by fitting the 1D spectrum with a Gaussian function. In the following, we adopt the redshift obtained from the 1D Gaussian fit ($z_\mathrm{\cii} = 3.39548 \pm 0.00007$). We compared this redshift estimate with that obtained from the VLT/SINFONI spectrum \citep[$z_\mathrm{opt} = 3.39535 \pm 0.00025$, ][]{Iani2021} and found that they agree within $1 \sigma$, increasing the reliability of the detection. 

The \cii\, emission of M2, the second image of our target, is also detected, although with lower S/N due to its distance from the center of the primary beam ($\sim 5\arcsec$) and, secondary, due to its lower magnification. We created 2D intensity maps of the \cii\, emission of M2 by averaging over the same channels that maximize the detection S/N of M1 (Figure \ref{fig:fig_maps}). We checked that by running instead our blind, iterative procedure to determine the channels that maximize the S/N of the \cii\, detection for M2 produces consistent results: the redshift of the emission changes by $\Delta z < 0.00005$ and the width of the line changes by $\Delta \mathrm{FWHM} < 15$ km s$^{-1}$. The \cii\, emission of M2 is detected with $\sim 6 \sigma$ significance.
To obtain a higher S/N spectrum, we stacked the 1D spectra of M1 and M2, extracted at the best-fitting position of the Fourier Transform of the 2D Gaussian model adopted for each galaxy image. The result is shown in Figure \ref{fig:fig_maps}.

We finally estimated the total \cii\, flux by fitting the channel-averaged emission line map of M1 in the \textit{uv} plane adopting the Fourier Transform of a 2D Gaussian model with the GILDAS task \texttt{uv\_fit}. Since the continuum was not detected, we did not subtract it. The best-fit model yields a \cii\, FWHM $\sim 0.8\arcsec \times 0.6\arcsec$, with an uncertainty of $\sim 0.1\arcsec$ on both axes. The total observed flux is $\mathrm{F_{\cii} = 7.5 \pm 0.7\, mJy}$ (over $\mathrm{\Delta v = 122\, km\, s^{-1}}$). This is the total \cii\, flux estimate (and best-fit model) that we use throughout the paper\footnote{We also checked whether a channel-by-channel fit would yield consistent results. In particular, we fit each channel encompassing the \cii\, emission with the Fourier Transform of a 2D Gaussian, with the center coordinates and the FWHM free to vary during the fit. We obtain a total observed \cii\, flux $\mathrm{F_{\cii} = 10.5 \pm 1.2\, mJy}$ over $\mathrm{\Delta v = 122\, km\, s^{-1}}$, consistent within $2\sigma$ with the previous estimate. This method allows for the recentering of the model in case the center and size of the emission changes with the frequency (or velocity). However, it might have the drawback of including positive noise peaks in the accounting of the flux. As a further check, we also fit the emission channel-by-channel with position and size of the model fixed to the values obtained by fitting the high S/N, channel-averaged map. In this case we estimate a flux $\mathrm{F_{\cii} = 8.4 \pm 0.7\, mJy}$, consistent within $\sim 1\sigma$ with both the other estimates.}.

When subtracting the best-fit model from the data and imaging the residuals (with natural weighting), we could detect some residual emission at the location of the NE clump. Despite the low significance of the detection ($\sim 3\sigma$), its overlap with the UV brightest clump suggests the possible presence of an additional unresolved component.

The relatively low significance of clumps in the channel-averaged map is likely due to the fact that they have a narrow \cii\, line (and likely a relatively small mass) implying a small contrast against a more diffuse and broader \cii\, emission (if present) and/or source confusion due to crowding. Indeed, when looking at the dynamics of our target, the emission from individual clumps peaks at slightly different velocity and is rather narrow ($\sim 40 - 80\, \mathrm{km\, s^{-1}}$, Section \ref{subsec:dynamics}). To assess if clumps are more prominent in narrower velocity ranges and more accurately determine their flux and significance, we searched for emission peaks in individual channel maps (Section \ref{subsec:channel_maps}).
We limit the morphological decomposition of the \cii\, emission to M1, due to the too low S/N of M2 that prevents a robust structural analysis.

\begin{figure*}[t!]
    \centering
    \includegraphics[width=0.9\textwidth]{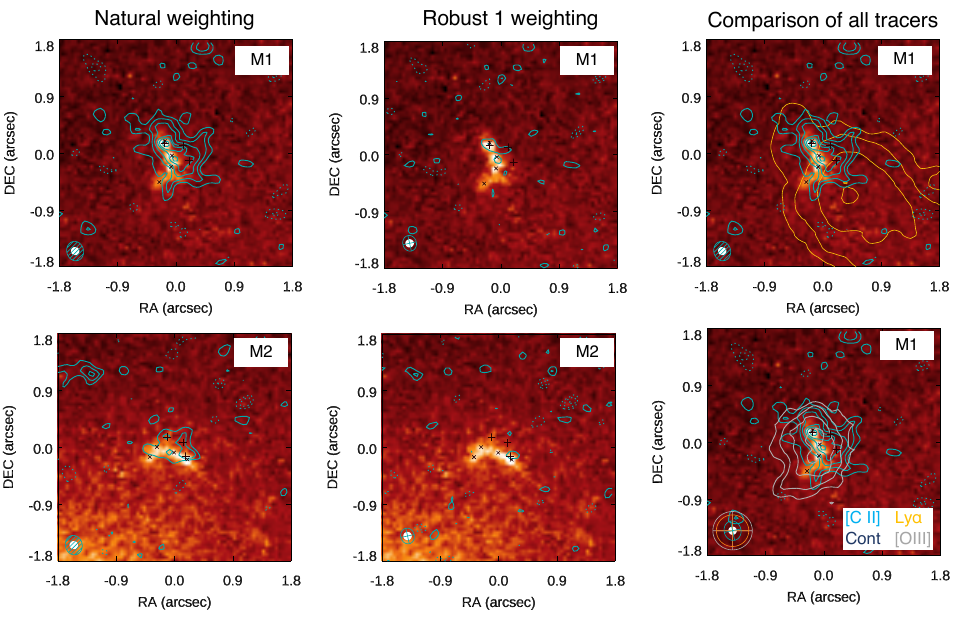}
    \caption{Comparison of UV and \cii\, morphology of our galaxy. We show both images covered by the primary beam, M1 and M2 (as indicated in the labels). \textit{Left panels}: \textit{HST} UV continuum data (background image) overlaid with the \cii\, data obtained with natural weighting (contours). The cyan solid and dashed contours indicate respectively positive and negative levels of 2, 3, 4, 5, 5.5, 6 rms. The crosses indicate the location of the clumps detected in \cii, while the x indicate the location of the clumps detected in UV continuum. The ALMA beam is reported as the cyan ellipse, while the \textit{HST} PSF is reported as the white-filled ellipse. \textit{Middle panels}: \textit{HST} UV continuum data (background image) overlaid with the \cii\, data obtained with GILDAS using robust weighting and a parameter of 1 (contours). Contour levels are the same as in the left panels. \textit{Right panels}: Comparison of the available tracers for the galaxy image M1: UV continuum (background), \cii\, (cyan contours as in previous stamps), \lya\, (orange contours with levels of 3, 5, 10, 18 rms, in the top panel), \oiii\, (gray contours with levels of 3, 5, 8, 11 rms in the bottom panel). The beams of each tracer are reported in the bottom right panel. All stamps have a size of $3.6\arcsec \times 3.6\arcsec$.}
    \label{fig:fig_comparison}
\end{figure*}

\subsection{Qualitative comparison of the \cii\, and UV morphology}
\label{subsec:qualitative_morphology}
Studying the morphology of clumpy galaxies using different tracers (e.g., rest-frame UV continuum, optical emission lines, IR emission lines and continuum) is key to unveiling the origin and nature of clumps. We aim to understand whether the \cii\, emission of our target is clumpy and if there are \cii\, clumps co-spatial with those detected at UV wavelengths.

To properly compare multiwavelength datasets we accurately calibrated the absolute astrometry of the \textit{HST} image. We selected 14 non-saturated and high S/N stars and compared their \textit{HST} sky-coordinates with the GAIA DR2 catalog \citep{Gaia_collaboration2016, Gaia_collaboration2018}. We matched the \textit{HST} astrometry to GAIA by applying the median offsets $\Delta \mathrm{RA} = 0.6600 \pm 0.0300$ and $\Delta \mathrm{Dec} = 0.0600 \pm 0.0500$. The absolute astrometry of the ALMA data, instead, is very accurate and there is no need to further refine it \citep{Farren2021}.

In Figure \ref{fig:fig_comparison}, we compare the morphology of the UV continuum with that of the channel-averaged 2D \cii\, map obtained as described in Section \ref{subsec:cii}. The peak of the \cii\, coincides with the brightest clump detected in the \textit{HST} image (in the following we refer to this clump as ``NE clump'', as it is located in the northeastern side of the galaxy). The \cii\, also shows a diffuse component largely overlapping with the UV continuum, despite being $\sim 1.8$ times more extended (see Section \ref{subsec:cii}). In particular, the \cii\, extends to a region where no UV is detected. A channel-by-channel analysis reveals the presence of a clump in that area (Section \ref{subsec:channel_maps} and \ref{subsec:dynamics}). In the following, we refer to this \cii-bright and UV-dark component as ``SW clump'' because it is located in the southwestern side of the galaxy. In between the NE clump and the SW one there is one more compact emission that is visible in the 2D \cii\, map and in the channel-by-channel analysis. It results in a tentative ($> 2.5\sigma$) detection over two consecutive channels (Section \ref{subsec:channel_maps}). In the following, we label this clump as ``central (C) clump''. We report it in the analysis below for completeness, although further observations are needed to confirm its detection.

Despite the lower S/N, also in M2 the \cii\, peak coincides with the brightest UV clump and the \cii\, emission extends to a region where the UV is not detected, as in the case of M1. This similarity and the symmetry of the \cii\, morphology of M1 and M2, which is expected given that the critical line passes in between these two galaxy images (Section \ref{subsec:lensing}), gives credit to the fact that the NE and SW clumps are actual star-forming regions rather than noise peaks (see also a more detailed discussion on this in Section \ref{subsec:significance}).

With the aim of increasing the spatial resolution of the observations, we also imaged the ALMA data using the "robust" (instead of "natural") weighting from GILDAS, with parameter 1, yielding a beam size of $\mathrm{FWHM} = 0.23\arcsec \times 0.19\arcsec$ and a noise r.m.s. of 0.23 mJy/beam, over a bandwidth of $\mathrm{102\, km s^{-1}}$, corresponding to the velocity width encompassed by the \cii\, emission (Section \ref{subsec:ALMA_data}). As expected, the diffuse \cii\, component is barely detected as it is resolved out, while the peak of the emission is still compact and detected in both M1 and M2. The C clump is tentatively detected in the 2D map of M1.

To assess whether the different morphology of the UV continuum and the \cii\, emission could arise from instrumental differences between optical and interferometric observations and/or observational biases, we used the \textsc{CASA} tasks \texttt{SIMOBSERVE} and \texttt{SIMANALYSE} to create mock ALMA \cii\, maps. We considered the best-fit parametric model obtained by fitting the UV image (Section \ref{subsec:uv_model}) and created mock \cii\, maps with the same angular resolution and r.m.s. as our actual ALMA observations. Figure \ref{fig:fig_mock} shows the mock maps of M1 imaged by using both natural and robust (with parameter 1) weighting. The morphology of the emission is clearly different from that of the actual \cii\, morphology: the peak of the emission coincides with the barycenter of the UV continuum and, when fitting it with a 2D elliptical Gaussian in the $uv$ plane, we recover sizes of $\mathrm{FWHM} \sim 0.4\arcsec \times 0.2\arcsec$, while the actual \cii\, emission has sizes of $\mathrm{FWHM} \sim 0.8\arcsec \times 0.6\arcsec$ (Section \ref{subsec:cii}). The NE clump is visible in the mock \cii\, maps, especially when robust weighting is used for the imaging. This suggests that the different morphology of \cii\, and UV emission are due to physical reasons rather than instrumental issues.

In Figure \ref{fig:fig_comparison} we show the morphology of the other available tracers, namely the \hb, \oiii, and \lya\, emission lines. While the optical lines (\hb\, and \oiii) are spatially coincident with the UV emission, the \lya\, instead is spatially offset by $\sim 0.16\arcsec \pm 0.02\arcsec$ that, at the redshift of our target, corresponds to $1.2 \pm 0.2$ kpc (de-lensed). A detailed analysis and comparison of the UV, \hb, \oiii, and \lya\, morphologies has already been presented by \cite{Iani2021}. For the present study we mainly focus on the morphology of the UV continuum from \textit{HST} and the \cii\, emission from ALMA.

\subsection{Identifying clumps in the \cii\, channel maps}
\label{subsec:channel_maps}

To identify clumps in individual channel maps, we followed an approach similar to the one developed by \cite{Dessauges-Zavadsky2019} and \cite{Dessauges-Zavadsky2023} to find GMCs in CO(4-3) observations of two lensed, $z \sim 1$ galaxies. We imaged individual channel maps using natural weighting and identified all the $\geq 4\sigma$ emissions within a radius corresponding to 1/3 of the primary beam. We then searched for all the spatially overlapping $>3\sigma$ emissions with the pre-identified emissions in at least two adjacent 20.3 km s$^{-1}$ channel maps. This allows us to exclude spurious noise peaks that are instead expected to be randomly distributed. Indeed, all the detections satisfying the above criteria are found to be co-spatial with the integrated \cii\, emission of M1. We considered as individual clumps those with $3 \sigma$ contours that are not spatially overlapping with each other in the same channel map or, if co-spatial, they are separated by at least one channel (i.e., they are not adjacent). In other words, different clumps could overlap spatially or in velocity, but not both. We identified three such clumps: the NE clump (spatially coincident with the brightest UV clump) is detected in 3 adjacent channels; the SW clump (not detected in the UV) is detected in 4 adjacent channels; finally clump C is detected in only 2 adjacent channels. We consider clump C as ``tentative'' as it was detected in only two channels. More observations with higher S/N are needed to confirm it.

\subsection{Measuring clumps \cii\, flux}
\label{subsec:flux_clumps}

We measured the flux of clumps both in the image plane and in the $uv$ plane and compared the results.
To measure fluxes in the image plane, we adopted customized apertures in each channel, such that they encompass all the emission above the local r.m.s. noise level. With this approach no aperture correction is needed. For each clump, the line-integrated fluxes were obtained by summing up the flux estimated in each adjacent channel \citep{Dessauges-Zavadsky2023}.

We also estimated the flux of each clump by fitting its emission in the $uv$ plane. When adopting Gaussian models, we obtain consistent results with those estimated from the image plane both in terms of flux and size (Table \ref{app:fluxes}). The FWHM of the best-fit Gaussians range between $0.26\arcsec$ and $0.60\arcsec$, corresponding to $800 - 1.9$ pc after magnification correction, in agreement with the FWHM radii we estimated from the 2D Gaussian fits performed in the channel maps encompassing the brightest \cii\, emission of each identified clump. The sizes of the best-fit Gaussians are 3 to 8 times larger than the sizes measured for the UV clumps of our target (Section \ref{subsec:uv_model}). They are also larger than typical GMC sizes measured using CO in lensed galaxies at $z \sim 1$ \citep{Dessauges-Zavadsky2019, Dessauges-Zavadsky2023}. We might be detecting kpc-scale \cii\, substructures, possibly due to the blending of smaller structures that are unresolved at the resolution of our observations. Alternatively, there might be some diffuse, intra-clump \cii\, emission where clumps are embedded, similarly to what is commonly seen in optical emission line maps (e.g., \citealt{ForsterSchreiber2011, Zanella2015, Zanella2019}). To investigate the second scenario, we also fit the clumps identified in channel maps with the Fourier Transform of 2D PSF models, representing compact clumps (radius $< 400$ pc) embedded in a more diffuse and broader emission. The center coordinates of the PSF models were left free during the fit (i.e., we did not fix them to the position of expected clumps). We obtain flux estimates for the SW, NE, and C clumps that are 1.2 - 1.5 times smaller than the Gaussian fit case (Table \ref{app:fluxes}). The discussion of whether to subtract possible intra-clump light or not when estimating the flux of clumps is a longstanding and unsolved issue at any wavelength (e.g., \citealt{ForsterSchreiber2011, Wuyts2012}) and goes beyond the scope of this work. Recent theoretical works suggest that most of the \cii\, is emitted in dense photodissociation regions associated with molecular clouds rather than in the diffuse, neutral medium due to the relatively high critical density of \cii\, ($\sim 3000$ cc) which is not achievable in the diffuse medium \citep{Pallottini2017b, vallini:2017, Olsen2017}. We decided to adopt the flux measurements obtained by fitting clumps with PSF models (Table \ref{tab:measurements}), as they are expect to be unresolved at the resolution of our observations (e.g., the size of the giant molecular clouds found in lensed galaxies at $z \sim 1$ is $R < 200\, \mathrm{pc}$, \citealt{Dessauges-Zavadsky2023}). We report the other flux estimates in the Appendix (Table \ref{app:fluxes}). Even when adopting fluxes obtained with the other reported methods, the conclusions of our work do not change substantially.
All fluxes have been corrected for lensing effects by considering the average magnification at the location of each clump.

\subsection{Reliability of clumps \cii\, detection}
\label{subsec:significance}

We performed additional tests to assess if the detected compact, spatially unresolved \cii\, components are physical entities (i.e., molecular clouds) or just peaks of correlated noise. In particular, we performed the following tests: we assessed whether point sources are needed to reproduce the radial amplitude of the \cii\, data; we computed the probability that the clumps are noise peaks given the symmetry of M1 and M2 offered by lensing; and we created mock \cii\, observations to understand what are the features that we expect to detect in actual \cii\, maps. We detail such tests in the following. 

\subsubsection{Radial amplitude analysis}
\label{subsubsec:radial_ampl}

\begin{figure}[t!]
    \centering
    \includegraphics[width=0.5\textwidth]{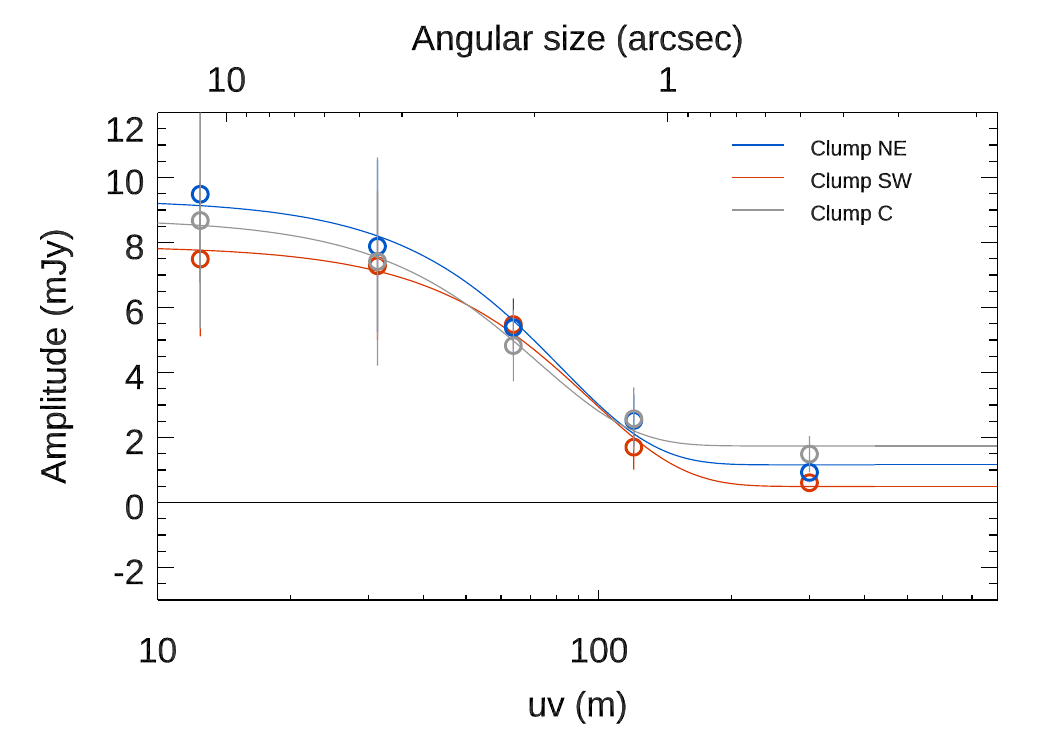}
    \caption{Signal amplitude as a function of the $uv$ distance, namely the baseline length. We extracted the signal amplitude at three different spatial locations, coinciding with the NE clump (blue), the SW clump (red), and the C clump (gray). We fit each dataset with a Gaussian model plus a constant. In the image plane these best-fit functions correspond  to a Gaussian with FWHM $\sim 2\arcsec$ and a PSF component. An additional constant, corresponding to another PSF component, is needed to fit the measurements.}
    \label{fig:radial_amplitude}
\end{figure}

We extract the amplitude of the signal as a function of the $uv$-distance, namely the baseline length, for the spectral channels spanning the \cii\, emission. For a point source, the amplitude is constant as a function of the $uv$ distance, while for an extended source the amplitude is maximum at short $uv$ distances and decreases at larger $uv$ distances. If the source is extended, we can determine its size and flux by fitting a 1D half-Gaussian profile to the radial amplitude profile. The physical size of the source is related to the FWHM of the Gaussian, while its total flux is the peak value of the Gaussian.
Figure \ref{fig:radial_amplitude} shows the radial amplitude of the \cii\, as a function of the $uv$ distance for M1, at the spatial position of the NE, SW, and C clumps. 

A Gaussian model with $\mathrm{FWHM \sim 2\arcsec}$ is needed to fit the data at short baselines, implying that the source is resolved. Its flux ranges between $\mathrm{F_{\cii,Gauss} \sim 8 - 10 \, mJy/\mu}$, depending on the extraction position, consistent with that estimated from the parametric modeling of the 2D surface brightness of the emission (Section \ref{subsec:cii}). We note that these are all observed estimates (i.e., not corrected for magnification).
However, an additional PSF (i.e., spatially unresolved) component is also needed to fit the longest baselines (i.e., shortest angular size) of the datasets extracted at the position of the clumps. This is shown by the fact that the amplitude of the signal remains constant at the longest $uv$ distances and does not go to zero. 
This supports the presence of unresolved \cii\, components in addition to the more extended emission.

\subsubsection{Probability of the \cii\, clumps being noise peaks}
\label{subsubec:probab}

Our target is a lensed galaxy and two images, M1 and M2, are inside the primary beam. For both images, the peak of the \cii\, emission coincides with the NE clump. In this Section, we assess whether the detected clump is an actual physical component of the galaxy or if it is an artifact, such as a noise peak, on top of an extended emission. To this aim, we calculated the probability of having a $> 3.5\, \sigma$ noise peak at a distance ranging between $0.35\arcsec$ and $0.45\arcsec$ from the barycenter, mimicking the compact \cii\, clump that is  detected in our dataset. In the primary beam, we injected at random positions a 2D Gaussian model with the same structural parameters obtained from the 2D fit of the \cii\, emission (Section \ref{subsec:cii}). With the CASA tasks \texttt{simobserve} and \texttt{simanalyze} we simulated 1000 mock \cii\, maps and in each of them used SExtractor to automatically detect noise peaks on top of the extended (Gaussian) emission. We then fit the detected peaks in the $uv$ plane with a point-like model. We computed the probability of finding noise peaks that have a $> 3.5\, \sigma$ significance and a distance from the Gaussian center in the range $0.35\arcsec - 0.45\arcsec$, consistent with the actual distance of our \cii\, clump from the galaxy barycenter (distance $\sim 0.4\arcsec$). We estimate a probability of 19\% of finding such a noise peak. However, our \cii\, is detected both in M1 and M2 and, in both cases, it coincides with the NE clump or, in other words, it has a specific position which is symmetric with respect to the critical line. We calculated the probability of having two noise peaks detected at $> 3.5\, \sigma$, within $0.35\arcsec - 0.45\arcsec$ from the galaxy barycenter, and with a symmetric position (e.g., being in the N-E side of the galaxy) to be 0.08\%. This shows that the compact \cii\, component overlapped with the NE clump is very likely to be a physical structure and not simply an observational artifact due to correlated noise or other instrumental effects. Analogous calculations hold also for the other \cii\, clumps which are tentatively detected both in M1 and M2 as elongations of the \cii\, emission in regions where no UV flux is detected (Figure \ref{fig:fig_maps}). The extremely low probability ($< 1\%$) of having a symmetric noise peak with respect to the critical line makes the detection of these \cii\, clumps reliable.

\begin{figure*}[t!]
    \centering
    \includegraphics[width=0.83\textwidth]{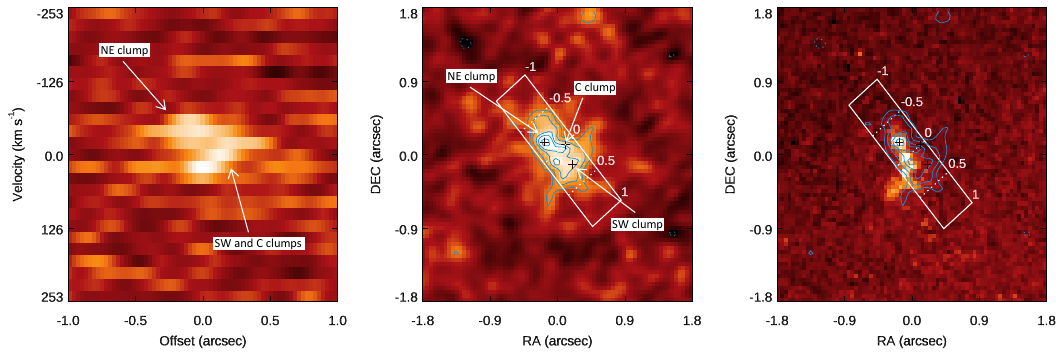}
    \includegraphics[width=0.82\textwidth]{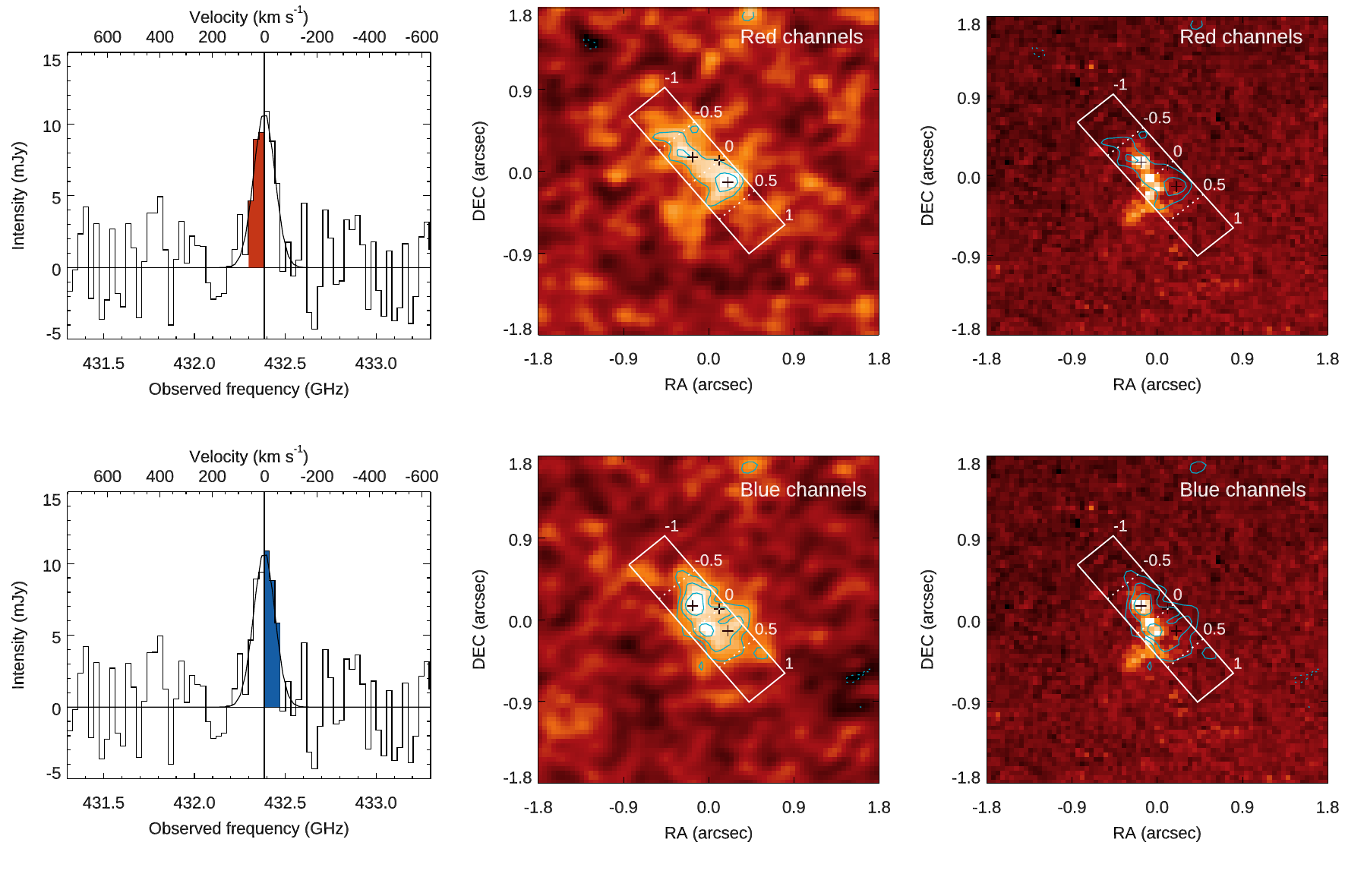}
    \includegraphics[width=0.83\textwidth]{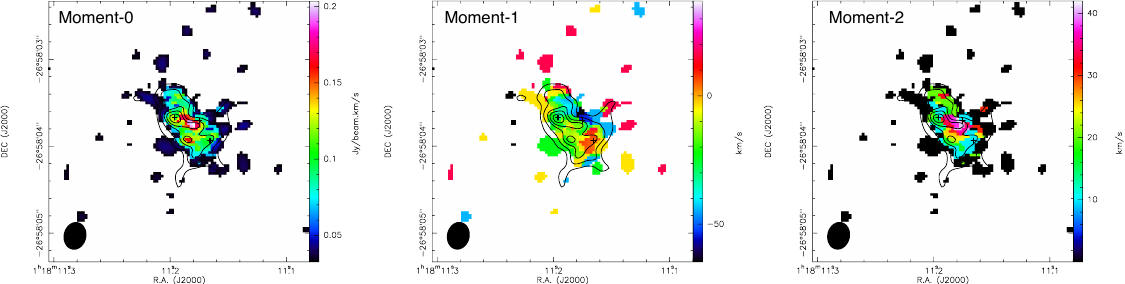}
    \caption{Dynamical properties of M1. \textit{First row, left panel}: position-velocity diagram of the \cii\, emission extracted along the slit shown in the next panels. \textit{First row, middle panel}: \cii\, map (background) overlaid with the \cii\, (3, 4, 5, 5.5, 6) rms contours (cyan) and the slit used to extract the PV diagram (white). The dashed lines mark offsets of $0.5\arcsec$ along the slit. \textit{First row, right panel}: UV continuum imaging (background) and \cii\, contours overlaid. \textit{Second row, left panel}: 1D spectrum showing the integrated \cii\, emission. In red we highlight the channels used to create the intensity map shown in the next panel. \textit{Second row, middle panel}: \cii\, intensity map obtained by averaging only the channels with positive velocity with respect to systemic (``red channels'') overlaid with (3, 4, 5) rms contours. \textit{Second row, right panel}: UV continuum (background), with \cii\, contours. Other symbols are as in previous panels. \textit{Third row, left panel}: 1D \cii\, spectrum, with channels used to produce the intensity maps of the next panel highlighted in blue. \textit{Third row, middle panel}: \cii\, intensity map obtained by averaging only the channels with negative velocity with respect to systemic (``blue channels'') with \cii\, (3, 4, 5) rms contours (cyan) overlaid. Other symbols are as in previous panels. \textit{Third row, right panel}: UV continuum (background) with \cii\, contours overlaid. Symbols are as in previous panels. \textit{Fourth row, right panel}: moment-0 map. \textit{Fourth row, middle panel}: moment-1 map. \textit{Fourth row, right panel}: moment-2 map. The contours are estimated from the intensity map and are the same as in Figure \ref{fig:fig_pv}. The black ellipse in the bottom left corner indicates the beam size. The crosses indicate the location of \cii\, clumps.}
    \label{fig:fig_pv}
\end{figure*}

\section{Results}
\label{sec:results}

In the following, we discuss the physical properties of our target galaxy and clumps as derived from the observables presented in previous sections.

\subsection{Star formation rate}
\label{subsec:SFR}
We estimated the star formation rate from the luminosity of the rest-frame UV continuum at 1500\AA. In particular, we considered the \textit{HST} imaging and modeled the 2D surface brightness profile of our target with GALFIT (Section \ref{subsec:uv_model}) to obtain the UV luminosity of individual clumps, as well as the total integrated one. We applied the recipes by \cite{Kennicutt1998}, after reporting them to a \cite{Chabrier2003} IMF, and obtained an estimate of the SFR. The galaxy has a total, unobscured, magnification-corrected star formation rate $\mathrm{SFR \sim 10 \pm 0.3\, M_\odot yr^{-1}}$ (Table \ref{tab:physical_prop}). The clumps have unobscured $\mathrm{SFR \sim 0.5 - 2\, M_\odot yr^{-1}}$ each, accounting in total for $\sim 5\% - 20\%$ of the UV light from the galaxy, as generally found both in lensed and non-lensed high-redshift star-forming regions \citep[e.g., ][]{Guo2018, Zanella2019, Mestric2022, Claeyssens2023}.

Our total, integrated SFR estimate is in agreement with that reported by \cite{Iani2021}, who estimated it from the luminosity of the H$\beta$ emission line extracted from the SINFONI 1D spectrum. They used the H$\beta$-to-SFR conversion factor by \cite{Kennicutt1998}, reported to a \cite{Chabrier2003} IMF, and obtained $\mathrm{SFR(H\beta) = 9.9 \pm 2.3\, M_\odot\, yr^{-1}}$. We did not correct the UV and H$\beta$ luminosity for dust extinction as the reddening that we estimated from the UV continuum $\beta$ slope is consistent with no dust extinction \citep{Iani2021}. This is also in agreement with the lack of dust continuum detection (Sections \ref{subsec:continuum} and \ref{subsec:lack_dust}). 
In Table \ref{tab:physical_prop} we also report the SFR and SFR surface density ($\Sigma_\mathrm{SFR}$) of individual clumps.

\subsection{Molecular gas mass}
\label{subsec:mgas}
We estimated the molecular gas mass of our target from the \cii\, luminosity. We considered a \cii\, luminosity-to-gas mass conversion factor $\mathrm{\alpha_{\cii}} = 31\, \mathrm{M_\odot/L_\odot}$ (with a scatter of 0.3 dex) as estimated by \cite{Zanella2018}. A comparable $\alpha_\mathrm{\cii}$ conversion factor has been estimated and used to derive molecular gas masses by other literature studies for galaxies at similar redshift \citep{Dessauges-Zavadsky2020, Gururajan2023, Bethermin2023} and in the local Universe \citep{Madden2020, Ramambason2023}, and it has been validated by theoretical arguments and simulations \citep{Pallottini2017a, Sommovigo2021, Vizgan2022}.

We obtain a total molecular gas mass for our galaxy, corrected for magnification, M$_\mathrm{H2} = (2.05 \pm 0.20) \times 10^9$ M$_\odot$ (Table \ref{tab:physical_prop}). This uncertainty accounts only for the errors related to the \cii\, luminosity. If we were to take into account also the uncertainties related to the $\alpha_\mathrm{\cii}$ conversion factor and magnification correction, the errors associated to M$_\mathrm{H2}$ would be $\sim 0.33$ dex larger. Such uncertainties do not account for systematic differences in the (unconstrained) \hi\, content of high-redshift galaxies. In fact, since \cii\, has a lower ionization potential than \hi\, (11.3 eV for \cii\, versus 13.6 eV for \hi), part of the \cii\, emission could arise from any gas phase (molecular, ionized, and atomic). However, it has been shown that the majority of the \cii\, luminosity ($\gtrsim 60 - 95$\%) is emitted by photodissociation regions (PDRs) and hence is associated with molecular gas, especially in the inner galaxy regions (e.g., \citealt{Stacey1991, Sargsyan2012, Pineda2013, Velusamy2014, Rigopoulou2014, Cormier2015, Croxall2017, Diaz-Santos2017}). This is supported also by simulations showing that $\sim 60 - 85$\% of the \cii\, is emitted by the molecular gas phase (e.g., \citealt{Vallini2015, Olsen2017, Accurso2017b}, although see \citealt{Heintz2021} for a discussion of \cii\, as an \hi\, tracer). Given that the \hi\, fraction is unconstrained at the redshift of our source and that we are studying the inner kpc of the galaxy, where the H$_2$ is expected to dominate over the \hi, we assume that the molecular gas is the main contributor to the \cii\, emission \citep{Combes2001}.
In Table \ref{tab:physical_prop} we report the $\mathrm{M_{H2}}$ and molecular gas mass surface density ($\Sigma_\mathrm{mol}$) of individual clumps.

\subsection{Dynamical properties}
\label{subsec:dynamics}
  We investigated the dynamics of our target by looking at the position-velocity (PV) diagram. PV diagrams are slices extracted from the data cube, along specific spatial directions. Figure \ref{fig:fig_pv} shows the PV diagram of the \cii\, emission extracted along the axis connecting the two detected \cii\, clumps. Distinct components are detected, separated by $\sim 0.3\arcsec$ and $\sim 50$ km s$^{-1}$. The component with negative velocity and spatial offset is mostly due to the NE clump, while the other more elongated emission is likely due to the contribution of several components (clump SW and clump C). To assess whether this is indeed the case, we created an emission map averaging the channels that are red-shifted with respect to the systemic \cii\, redshift and a second map averaging the blue-shifted channels (Figure \ref{fig:fig_pv}). The first map (red-shifted channels) shows mainly the SW clump (positive offset with respect to the galaxy barycenter) plus a ``bridge'' emission connecting it with the NE clump location. The second map (blue-shifted channels) instead shows the NE clump and the region around the galaxy barycenter. These findings are in agreement with the results obtained by analyzing individual channel maps (Section \ref{subsec:channel_maps}) and support the result that the \cii\, emission is made of multiple, distinct components. 

  Finally, we created moment maps with the $\texttt{immoments}$ task of the CASA software. We adopted a $2.5\sigma$ threshold above the r.m.s. noise level. The moment-0 (intensity), moment-1 (velocity), and moment-2 (velocity dispersion) maps are shown in Figure \ref{fig:fig_pv}. They are not corrected for beam-smearing, instrumental line spread function, nor lensing effects. In the moment-0 map the SW clump appears more clearly than in the intensity map, likely due to its relatively large velocity dispersion (see also the moment-2 map), while the NE clump is not prominent in the moment-0 map due to its narrow velocity width. The moment-1 map does not show ordered rotation. The velocity dispersion ranges between 10 and 35 km s$^{-1}$, it peaks on the SW clump and reaches its minimum in the external regions of the galaxy, which are also less affected by beam smearing. This seems to be a dispersion-dominated galaxy and similar results are found when considering the moment maps reconstructed in the source plane (Figure \ref{fig:moment_maps}). The galaxy appears dispersion-dominated also when the \oiii 5007\AA\, line emission is considered, as reported by \citet[although the \oiii\, observations are seeing-limited and hence have coarser angular resolution than the \cii\, data used in our study]{Livermore2015}. The dispersion-dominated nature of the galaxy disfavors the formation of the observed clumps due to violent disk instability in an isolated, rotationally supported disk \citep[e.g., ][]{Dekel2014, Bournaud2014, Ceverino2015, Tamburello2015}. The kinematics of the galaxy rather suggests that these clumps formed as a consequence of a recent interaction, although we do not detect companion galaxies within the primary beam (up to $\sim 100\, \mathrm{kpc}$ distance from our target). Alternatively, one of the clumps might be an ex situ object, such as a low-mass galaxy, merging with our target and giving rise to both its dispersion-dominated dynamics and the formation of clumps due to gas instability and fragmentation \citep{Teyssier2010, Bournaud2011, Renaud2014, Calabro2019, Zanella2019}.  
  In the following, the contribution of rotation to the observed velocity dispersion is neglected. However, more robust dynamical modeling, which is beyond the scope of this paper, is needed to confirm and refine these findings.

\subsection{Expected size of clumps from gravitational instability}
 \label{subsec:size}
  
  We investigated whether we should indeed expect to detect clumps given the resolution of our observations, the gas mass surface density of our target galaxy, and its velocity dispersion. Molecular clouds that form in situ due to the gas gravitational instability, are expected to have an average size that is comparable to the Jeans length:
  \begin{equation}
      \lambda_\mathrm{J} = \frac{\sigma^2}{2 \pi G \Sigma_\mathrm{gas}}
      \label{eq:jeans}
  \end{equation}

  where G is the gravitational constant, $\Sigma_\mathrm{gas}$ is the gas surface density, and $\sigma$ is the gas velocity dispersion within the disk \citep{Toomre1964, Elmegreen2009a, Elmegreen2009b, Canameras2017, Gullberg2018}. We estimated the magnification-corrected gas surface density of the galaxy $\Sigma_\mathrm{gas} = 525\, \mathrm{M_\odot\, pc^{-2}}$ by considering its molecular gas mass as estimated from the \cii\, luminosity (Section \ref{subsec:mgas}) and the size of the extended \cii\, emission (Section \ref{subsec:cii}). From the moment-2 map (Figure \ref{fig:fig_pv}) we measure velocity dispersions in the range $\sigma \sim 10 - 35$ km s$^{-1}$, where the smallest velocity dispersion is found in the outskirts of the galaxy which is less affected by beam smearing. The integrated velocity dispersion of the \cii\, line is $\sigma \sim 37$ km s$^{-1}$ and once we correct it for the instrumental broadening we obtain an intrinsic velocity dispersion $\mathrm{\sigma_{int} \sim 20\, km\, s^{-1}}$. By adopting an average velocity dispersion $\mathrm{\sigma \sim 20\, km\, s^{-1}}$, we obtain gravitationally unstable scales $\mathrm{\lambda_J \sim 100\, pc}$ which are compatible with the fact that our \cii\, clumps are unresolved (i.e., appear as point-like sources) at the resolution of our observations (beam $\mathrm{FWHM = 0.26\arcsec \times 0.31\arcsec}$ that corresponds to $\mathrm{R_e \lesssim 400\, pc}$ after correcting for lensing magnification). Turbulence and gravitational collapse are expected to trigger the generation of stars inside these parent molecular clouds \citep{Elmegreen2009b}. The clumps  are characterized by  high density ($\Sigma_\mathrm{gas} > 10^3\, \mathrm{M_\odot pc^{-2}}$) and high  gravitational pressure, $\mathrm{P_{grav}/k_b \sim G \Sigma_{gas}^2 > 10^7\, K cm^{-3}}$.
Such  extreme pressure values are typical of the most highly pressurized medium in molecular clouds of star-forming galaxies \citep{Elmegreen1997, Calura2022}.  This scenario is broadly consistent with the fact that we detect  clumps in the UV continuum images, with effective radii $\mathrm{R_e \lesssim 250\, pc}$ (Table \ref{tab:physical_prop}).

\begin{figure*}[t!]
    \centering
    \includegraphics[width=0.9\textwidth]{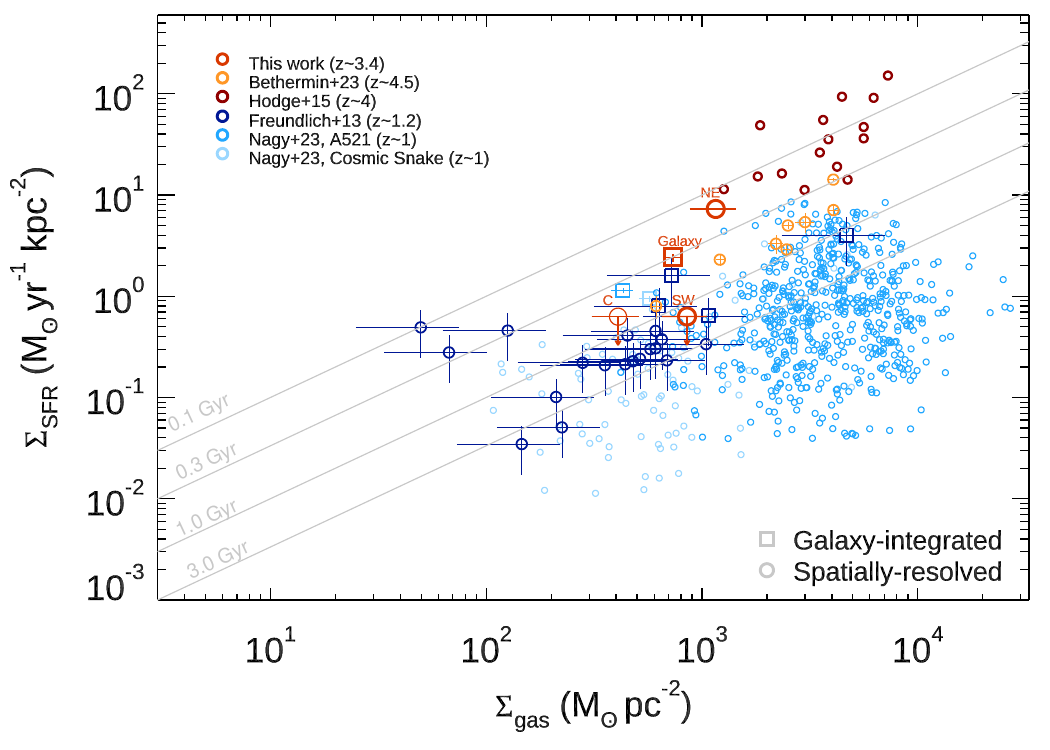}
    \caption{Clumps location in the Schmidt-Kennicutt plane. We compare our findings (red symbols), with available subkiloparsec observations from the literature at $z \sim 1 - 4$ (colored symbols, \citealt{Freundlich2013, Hodge2015, Nagy2023, Bethermin2023}). Galaxy-integrated measurements are reported as squares, while spatially resolved measurements as circles. Clump C, which is considered as tentative, is indicated with the red, thinner, circle. We also indicate lines of constant depletion time (gray lines).} 
    \label{fig:SFE}
\end{figure*}

\section{Discussion}
\label{sec:discussion}

  \subsection{Gas depletion time and Schmidt-Kennicutt plane}
  \label{subsec:SK}

  By comparing the SFR and molecular gas mass surface densities of galaxies and clumps in the traditional Schmidt-Kennicutt (SK) plane \citep{Schmidt1959, Kennicutt1998}, it is possible to assess their depletion time $t\mathrm{_{dep} \sim \Sigma_{mol}/\Sigma_{SFR} \sim M_{H2}/SFR}$. This ratio indicates the time that is needed to turn the molecular gas reservoir into stars, assuming a constant SFR. The molecular gas and young stars are expected to be co-spatial at the scale of individual GMCs if the star formation happens in quasi-equilibrium for several dynamical times. On the contrary, if the star formation process happens quickly and GMCs are disrupted soon after the formation of the first massive stars, then molecular gas and young stars are decorrelated at the small scales \citep{Schruba2010, Kruijssen2019, Kim2022, Nagy2023}. Spatially resolved studies are crucial to constrain what is the depletion time at subkiloparsec scales and unveil the possible (de)correlation between star-forming regions and GMCs. Figure \ref{fig:SFE} shows the location of our galaxy and clumps in the  $\Sigma_\mathrm{mol} - \Sigma_\mathrm{SFR}$ plane, and compares it with literature galaxies and their subkiloparsec regions \citep{Freundlich2013, Hodge2015, Nagy2023, Bethermin2023}. We only included galaxies at $z \sim 1 - 4$ and their subkiloparsec regions that are detected at sub-mm wavelengths. In Figure \ref{fig:SFE} we show the SFR and $\mathrm{M_{mol}}$ surface densities averaged over an area of $\sim 400\, \mathrm{pc}$ radius, which is the physical scale corresponding to the angular resolution of the observations. This is consistent with the area considered to estimate surface densities of sub-galactic regions in the literature \citep{Freundlich2013, Hodge2015, Nagy2023, Bethermin2023}. 
  Adopting smaller sizes would not change our results (e.g., depletion time estimates), as the location of clumps in the SK plane would change along the curves of iso-depletion time.
    Our \cii-detected clumps span a large range of depletion time. The NE clump, which is detected both in UV and \cii, has a very short $t_\mathrm{dep} = 0.16 \pm 0.04\, \mathrm{Gyr}$. On the contrary, the SW and C clumps, which are not detected in the UV, have a rather long $t_\mathrm{dep} > 0.65\, \mathrm{Gyr}$.
    We compare our results with the depletion time of individual sub-galactic regions in apertures of $\sim 400$ pc in radius from two $z \sim 1$, strongly lensed galaxies: the Cosmic Snake and A521 \citep{Nagy2023}\footnote{We rescaled the SFR from \cite{Nagy2023} to a \cite{Chabrier2003} IMF, to properly compare with the SFR estimated for our clumps.}. 
    The apertures of $\sim 400$\, pc in radius (in the source plane) are comparable to the size upper limit of our clumps (Table \ref{tab:physical_prop}). 
    The depletion time in such regions is on average $t_\mathrm{dep} > 1\, \mathrm{Gyr}$, although a large scatter is observed (Figure \ref{fig:SFE}). One of the reasons reported by \cite{Nagy2023} to explain the scatter of the SK at subkiloparsec scales is the fact that the $t_\mathrm{dep}$ measurements largely depend on whether they are performed in apertures dominated by star-forming regions (i.e., UV peaks) or GMCs (i.e., CO peaks). Randomly selected regions might have a large range of $t_\mathrm{dep}$, depending on whether they capture only one phase (star-forming vs gaseous) or both. The average depletion time observed in the Cosmic Snake and A521 is in good agreement with the $t_\mathrm{dep}$ upper limits estimated for our SW and C clumps. The fact that no UV continuum is observed at the location of the \cii\, emission of these regions, implying a rather long gas depletion timescale, might indicate that we are witnessing the onset of star formation, when the molecular gas reservoir has not been consumed or disrupted yet and the star formation is still embedded. The fact that no SFR is detected in these regions might also be due to the fact that the UV continuum is tracing stellar populations with ages $\sim 100$\, Myr, which might not be present yet.  Observations of tracers probing earlier phases of star formation (e.g., H$\alpha$ emission probing ages $\lesssim 10$\, Myr) are needed to assess whether young stellar populations are present in these regions. On the contrary, the NE clump has a $\sim 4\times$ shorter depletion time than the average $t_\mathrm{dep}$ of subkiloparsec regions in the Cosmic Snake and A521, rather comparable to sub-galactic regions found in starbursts \citep{Hodge2015} and $z \sim 4$ galaxies \citep{Bethermin2023}. Other spatially resolved studies of high-redshift galaxies \citep{Rawle2014, Chen2017, Calura2021} showed regions with short depletion time $t_\mathrm{dep} \sim 100\, \mathrm{Myr}$, although we do not include them in Figure \ref{fig:SFE} as they encompass larger sizes ($\gtrsim 1\, \mathrm{kpc}$) than our observations and/or have higher redshift ($z \gtrsim 5$). One possible explanation for the short depletion time of the NE clump is the fact that an intense and rather long-lasting episode of star formation is ongoing in this region, possibly implying molecular gas replenishment from the surrounding regions to sustain star formation over $\sim 100$\, Myr timescale (the time span probed by the UV continuum). Another possibility is the fact that the UV and \cii\, emission are spatially decorrelated over scales $< 300 - 400$\, pc, which are not resolved with our observations, and therefore they appear co-spatial because of the lack of angular resolution. Observations with resolution $\sim 0.05\arcsec - 0.1\arcsec$ (corresponding to $\lesssim 150$\, pc in the source plane) are needed to shed light on this.
  
\subsection{The lack of dust continuum}
\label{subsec:lack_dust}

Besides the \cii\, line emission, our ALMA observations also probe the dust continuum at wavelength $\lambda_\mathrm{obs} \sim 700\, \mathrm{\mu m}$ (corresponding to rest-frame $\lambda_\mathrm{rf} \sim 160\, \mathrm{\mu m}$). The continuum is not detected in the 1D spectrum nor in the 2D maps (Section \ref{subsec:continuum})\footnote{The continuum map (Figure \ref{fig:fig_maps}) shows a $\mathrm{S/N \sim 3.1}$ peak offset by $\sim 0.26"$ (corresponding to $\sim 800\, \mathrm{pc}$ on the source plane) from the \cii\, SW clump. However, due to the low significance, the offset from the \cii\, detection, and the fact that the same feature is not present in the second galaxy image (M2), we regard it as a noise peak and consider the SW clump undetected in the continuum.}. We estimated a $5\sigma$ continuum upper limit for M1 by fitting a Gaussian model or a PSF model in the \textit{uv} plane (see Section \ref{subsec:continuum}). We corrected both upper limits for lensing effects, by applying an average magnification correction. We obtain a continuum flux 5$\sigma$ upper limit of $\mathrm{F_{cont} < 105\, \mu Jy}$ when considering a Gaussian model and $\mathrm{F_{cont} < 34\, \mu Jy}$ when considering a PSF model. These are more stringent upper limits than most values reported in the literature for galaxies at similar redshift. \cite{Koprowski2020} targeted the $870\, \mu$m dust continuum for a sample of 250 Lyman Break Galaxies (LBGs) at $z \sim 3$ and obtained 41 detections. By stacking the non-detections, they obtained a tentative $3.2 \sigma$ detection with a total flux of $\mathrm{F_{870\mu m} = 65 \pm 20 \, \mu Jy}$. Similar results are obtained by \cite{Coppin2015} that, by stacking the $850\, \mu$m continuum data of LBGs at $z \sim 3$ and $z \sim 4$ obtained average fluxes $\mathrm{F_{850\mu m} = 250 \pm 30 \, \mu Jy}$ at $z \sim 3$ and $\mathrm{F_{850\mu m} = 410 \pm 30 \, \mu Jy}$ at $z \sim 4$. The $850\mu$m dust content of main-sequence $z \sim 4.5-6$ galaxies was also investigated by the ALPINE survey \citep{Bethermin2020}: 23 target galaxies out of 118 were detected with $\mathrm{F_{850\mu m} \gtrsim 115\, \mu Jy}$. Non detections were stacked in bins of \cii\, luminosity and yielded average continuum fluxes $\mathrm{<F_{850\mu m}> \sim  50 - 300\, \mu Jy}$ \citep{Bethermin2020}, approaching the upper limit of our target. Finally, \cite{Capak2015} analyzed individual LBGs at $z \sim 5$ and estimated from their non-detections a 3$\sigma$ upper limit  $\mathrm{F_{850\mu m} \lesssim 30\, \mu Jy}$, comparable to ours. Hence, lensing magnification allowed us to place an upper limit on the continuum flux which is more stringent than most of those estimated for (non-lensed) galaxies at similar redshift. Our result is consistent with the blue slope of the UV continuum $\beta = -2.53 \pm 0.15$ and the reddening $\mathrm{E(B-V)_{cont} < 0.16}$ mag estimated from the optical data \citep{Iani2021}, indicating that star formation is mostly unobscured in our target. 

However the question remains about the SW clump that we detected in \cii, but not at UV or IR continuum wavelengths. We estimated the total IR luminosity ($\mathrm{L_{IR}}$) of the SW clump by multiplying the $158\mu$m continuum upper limit of the SW clump by the ratio between the monochromatic continuum luminosity ($\mathrm{\nu L_\nu}$) and the $\mathrm{L_{IR}}$ at the rest-frame wavelength associated with the \cii\, line emission ($\mathrm{\nu L_\nu}/\mathrm{L_{IR}} = 0.133$), as reported by \cite{Bethermin2017, Bethermin2020} for galaxies at similar redshift. We obtain an upper limit on the total IR luminosity $\mathrm{L_{IR} < 5\times 10^{8}\, L_\odot}$. Finally, by adopting the $\mathrm{L_{IR}}$-to-SFR conversion proposed by \cite{Kennicutt1998}, converted to a \cite{Chabrier2003} IMF, we placed an upper limit on the obscured SFR of the SW clump $\mathrm{SFR \lesssim 0.1\, M_\odot yr^{-1}}$. This is consistent with the 3$\sigma$ unobscured SFR upper limit that we obtained from the UV from the SW clump ($\mathrm{SFR < 0.3\, M_\odot\, yr^{-1}}$) and with the typical SFR of clumps at these redshifts which is in the range $\mathrm{SFR \sim 0.01 - 10\, M_\odot yr^{-1}}$ \citep[e.g., ][]{Guo2018, Zanella2019, Mestric2022, Claeyssens2023}. Finally, from the \cii\, luminosity of the SW clump into SFR, adopting the relation by \cite{deLooze2014} for starbursts, we obtain $\mathrm{SFR_{\cii} = 0.3 \pm 0.1}$. If we were to adopt the SFR-to-\cii\, conversion factor estimated for high-redshift galaxies instead \citep{deLooze2014}, we would obtain similar results ($\mathrm{SFR_{\cii} = 0.2 \pm 0.1}$), consistent with the estimate derived from the UV and with the non-detection in the ALMA continuum map. This suggests that greater sensitivity (hence deeper observations and/or larger magnification) is needed to detect the IR continuum emission of individual clumps. 

It has been suggested that the clumps ubiquitously detected at UV and optical wavelengths could be an effect of dust inhomogeneities across the galaxy disk that make the UV light appear patchy, enhancing some small dust-free structures while hiding the most attenuated ones \citep{Buck2017}. While attenuation could indeed play an important role in dusty submillimeter galaxies \citep{Hodge2016, Rujopakarn2016, Ivison2020}, in lower mass, dust-poor galaxies such as our target, the physical properties of individual clumps seem to be mostly unaffected by dust. The fact that the NE clump of our target is also detected in \cii\, which, unlike the UV, is not strongly affected by dust, supports the scenario in which clumps in low-mass galaxies are actual physical structures. Finally, the non-detection of the far-IR continuum emission implies that the offset between the Ly$\alpha$ and UV continuum emission (Section \ref{subsec:target} and Figure \ref{fig:fig_comparison}) in this galaxy is likely caused by scattering due to the presence of neutral gas rather than dust.

\section{Summary and conclusions}
\label{sec:conclusions}

We investigated the \cii\, emission of a clumpy galaxy at redshift $z \sim 3.4$ lensed by the foreground galaxy cluster Abell2895. Our ALMA data cover two multiple images of the target, M1 and M2. Additional ancillary data from \textit{HST} (probing the rest-frame UV continuum), VLT/MUSE (probing the Ly$\alpha$ emission), and VLT/SINFONI (probing the \oiii\, and \hb\, emissions) are available. We found that:

\begin{itemize}
\item The spatially integrated \cii\, emission is detected at $\sim 10\sigma$ significance. It mostly overlaps with the rest-frame UV continuum emission from \textit{HST}, but it is $\sim 1.8$ times spatially more extended. 
\item We detected the \cii\, emission of individual clumps that we labeled NE, SW, and C. The NE clump coincides with the brightest clump detected in the UV continuum, while the SW and C clumps are not detected in the available \textit{HST} imaging. Both images, M1 and M2, show comparable morphology, although M2 is observed with lower S/N due to the fact that it is located $\sim 5\arcsec$ away from the center of the ALMA primary beam (and secondary it also has slightly lower magnification).
\item Our observations do not resolve the \cii\, clumps, yielding intrinsic (magnification-corrected) radii $\mathrm{R_e \lesssim 300 - 400\, pc}$. The fact that they are not resolved is in agreement with expectations for molecular clouds formed by fragmentation due to gravitational instabilities, yielding Jeans unstable scales $\mathrm{\lambda_J \sim 100\, pc}$, given the velocity dispersion and the molecular gas surface density of our target.
\item The galaxy is dispersion-dominated as shown by the position-velocity diagram and moment maps. This indicates that the formation of clumps likely did not occur due to gravitational disk instability in an isolated disk, but it is rather induced by a merger. We did not detect any galaxy interacting with our target within $\sim 100\, \mathrm{kpc}$, although one of the clumps might actually have an ex situ origin and be the remnant of the merging satellite.
\item We estimated the molecular gas mass of individual clumps from their \cii\, luminosity, adopting the conversion factor proposed by \citet[but see also \citealt{Dessauges-Zavadsky2020, Madden2020}]{Zanella2018}. We find $\mathrm{M_{mol} \sim 10^8\, M_\odot}$. The NE clump, which is detected both in \cii\, and UV, has a short depletion time $t_\mathrm{dep} = 0.16 \pm 0.04\, \mathrm{Gyr}$, comparable with sub-galactic regions in high-redshift galaxies \citep{Hodge2015, Bethermin2023}. The SW and C clumps instead have longer depletion time ($t_\mathrm{dep} > 0.65\, \mathrm{Gyr}$), similar to $z \sim 1$ sub-galactic regions \citep{Freundlich2013, Nagy2023}.
\item We do not detect the dust continuum down to $\mathrm{F_{cont} < 34\, \mu Jy}$. This is consistent with the blue UV continuum slope ($\beta \sim -2.53 \pm 0.15$) estimated from the VLT/MUSE data \citep{Iani2021}. The continuum non-detection is consistent with the SFR upper limit derived from the UV ($\mathrm{SFR_{UV} < 0.3\, M_\odot yr^{-1}}$) and from the \cii\, when adopting the \cite{deLooze2014} conversion ($\mathrm{SFR_{\cii} < 0.3\, M_\odot yr^{-1}}$). Deeper observations are needed to detect the dust continuum of individual clumps. This also suggests that in low-mass galaxies such as our target, clump detection is not significantly affected by dust distribution and inhomogeneities \citep{Buck2017}.
\end{itemize}

Exquisite spatial resolution and sensitivity are needed to detect clumps and their parent molecular clouds. Most high-redshift studies aiming to detect the dust continuum, CO, or \cii\, emission of individual clumps yielded tentative or non-detections \citep{Cibinel2015, Gullberg2015, Hodge2016, Rujopakarn2019, Ivison2020, Calura2021}. This might be due to a number of reasons: the relatively coarse spatial resolution of the observations and/or lack of sensitivity (e.g., \citealt{Cibinel2015, Calura2021}), the short depletion time of molecular clouds, and/or the fact that the targets were submillimeter galaxies that are bright in the IR, but not observed at UV wavelengths, hence it is unknown whether they host UV clumps in the first place \citep{Gullberg2018, Hodge2016, Rujopakarn2019, Ivison2020}. Gravitational lensing allowed \cite{Dessauges-Zavadsky2019} and \cite{Dessauges-Zavadsky2023} to achieve greater sensitivity and spatial resolution and detect the CO(4-3) emission from individual molecular clouds hosted by two main-sequence, clumpy galaxies at $z \sim 1$. Our work extends the detection of clumps at sub-mm wavelengths up to $z \sim 3.4$ and suggests that \cii\, is a promising tracer of molecular clouds at high redshift. Larger samples of lensed clumps observed with \cii\, will be needed to further strengthen our conclusions. The simultaneous availability of other tracers (e.g., dust-continuum and/or CO emission) will allow us to constrain further the physical properties of molecular clouds down to scales of hundreds of pc and pinpoint the initial conditions for clumps formation.

\section*{Acknowledgements}

We thank the referee whose comments and suggestions helped us to improve the clarity of the paper.
The research activities described in this paper have been co-funded by the European Union – NextGenerationEU within PRIN 2022 project n.20229YBSAN - Globular clusters in cosmological simulations and in lensed fields: from their birth to the present epoch. AF and MH acknowledge support from the ERC Advanced Grant INTERSTELLAR H2020/740120. C.-C.C. acknowledges support from the National Science and Technology Council of Taiwan (NSTC 111-2112M-001-045-MY3), as well as Academia Sinica through the Career Development Award (AS-CDA-112-M02). EI acknowledges funding from the Netherlands Research School for Astronomy (NOVA).
 This paper makes use of the following ALMA data: 2019.1.01676.S. ALMA is a partnership of ESO (representing its member states), NSF (USA), and NINS (Japan), together with NRC (Canada), MOST and ASIAA (Taiwan), and KASI (Republic of Korea), in cooperation with the Republic of Chile. The Joint ALMA Observatory is operated by ESO, AUI/NRAO and NAOJ.

\section*{Data Availability}

The data used in this study are publicly available from telescope archives. Software and derived data generated for this research can be made available upon reasonable request made to the corresponding author.

%
%

\bibliographystyle{aa}
\bibliography{bibliography} 

\begin{appendix}

\section{Mock \cii\, maps}

\begin{figure*}[t!]
    \centering
    \includegraphics[width=0.7\textwidth]{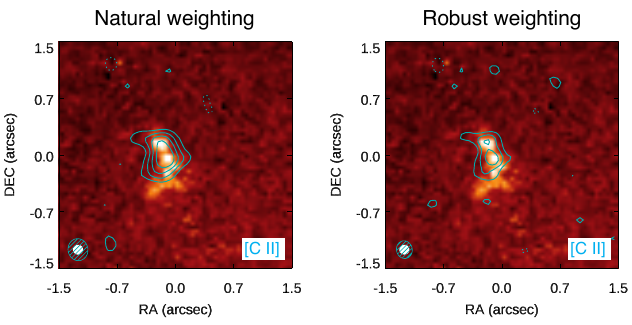}
    \caption{Comparison of UV and the \cii\, morphology obtained by simulating mock ALMA data for the galaxy image M1. \textit{Left panel}: \textit{HST} UV continuum data (background image) overlaid with the \cii\, data obtained with natural weighting (contours). The cyan solid and dashed contours indicate respectively positive and negative levels of 3, 5, 7, 9 rms. The ALMA beam is reported as the cyan ellipse, while the \textit{HST} PSF is reported as the white filled ellipse.  \textit{Right panel}: \textit{HST} UV continuum data (background image) overlaid with the \cii\, data obtained with robust weighting (contours). Contour levels are the same as in the left panel. Each stamp has a size of $3\arcsec \times 3\arcsec$.}
    \label{fig:fig_mock}
\end{figure*}

To check whether the \cii\, and UV light distribution is intrinsically different and not an observational effect, we created mock \cii\, maps with the \textsc{CASA} tasks \texttt{SIMOBSERVE} and \texttt{SINANALYSE}. We considered the best-fit \textsc{GALFIT} model obtained by fitting the UV continuum. We simulated the same ALMA configuration and integration time used for the actual observation. In Figure \ref{fig:fig_mock} we show the \cii\, mock maps imaged with natural and robust (with parameter 1) weights. In both cases imaging is performed with GILDAS.

\section{Moment maps on the source plane}
We created moment maps on the source plane, using the lensing model described in Section \ref{subsec:lensing}. To create the intensity (moment-0), velocity (moment-1), and velocity dispersion (moment-2) maps, we used the same procedure adopted for the image plane moment maps and described in Section \ref{subsec:dynamics}.

\begin{figure*}[t!]
    \centering
    \includegraphics[width=\textwidth]{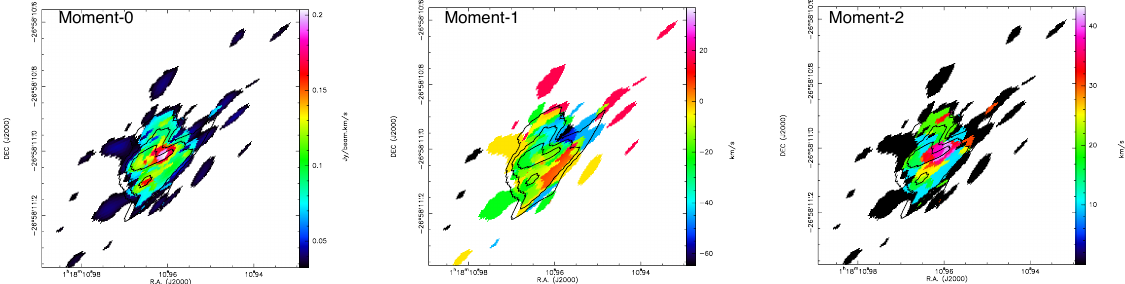}
    \caption{Moment maps of our target based on the source plane reconstruction. The symbols are the same as in Figure \ref{fig:fig_pv}.}
    \label{fig:moment_maps}
\end{figure*}

\section{Estimates of the \cii\, flux of clumps}
We estimated the flux of clumps detected in individual channel maps (Section \ref{subsec:channel_maps}) using different approaches (Section \ref{subsec:flux_clumps}). In the following, we briefly describe each method and we report the related measurements in Table \ref{app:fluxes}.
\begin{enumerate}
    \item \textbf{Image plane.} We fit the \cii\, emission in the image plane using customized apertures in each channel, such that they include all the emission above the local r.m.s. noise level. For each clump, the line-integrated fluxes were obtained by summing up the flux estimated in each adjacent channel.
    
    \item \textbf{\textit{uv} plane, Gaussian.} We fit the \cii\, in the $uv$ plane using the Fourier Transform of elliptical Gaussian 2D models, in each channel encompassing the emission. The center coordinates, FWHM, and position angle of the Gaussian were free to change in each channel. The total flux was estimated by summing up the fluxes obtained in each adjacent channel.
    
    \item \textbf{\textit{uv} plane, PSF free.} We fit the emission in the $uv$ plane adopting the Fourier Transform of a 2D PSF model, in each channel. The center coordinates were free to change in each channel. The total flux was estimated by summing up the fluxes obtained in each adjacent channel.
    
    \item \textbf{\textit{uv} plane, PSF fixed.} We fit the emission in the $uv$ plane with the Fourier Transform of a 2D PSF model, whose coordinates were fixed between different channels (i.e., forced to remain the same in each channel). The total flux was estimated by summing up the fluxes obtained in each adjacent channel.

    \item \textbf{\textit{uv} plane, PSF fit average.} We fit the average \cii\, map in the $uv$ plane adopting the Fourier Transform of a 2D PSF model. The maps were obtained by averaging in the $uv$ plane (with the GILDAS task \texttt{uv\_average}) only the channels encompassing the emission of a given clump.
\end{enumerate}
The fluxes reported in the main text have been estimated adopting method 3.

\begin{table*}
    \centering
    \caption{Observed (i.e., not corrected for magnification) \cii\, flux measurements of clumps.}
    \begin{tabular}{c c c c c c c}
    \toprule
    \midrule
    ID & $\mathrm{\mu F_{\cii}}$ & $\mathrm{\mu F_{\cii}}$ & $\mathrm{\mu F_{\cii}}$ & $\mathrm{\mu F_{\cii}}$ & $\mathrm{\mu F_{\cii}}$ & $\Delta$ v \\
       & Image plane & $uv$ plane, Gaussian & $uv$ plane, PSF free & $uv$ plane, PSF fixed & PSF fit average & \\
       & (mJy km s$^{-1}$) & (mJy km s$^{-1}$) & (mJy km s$^{-1}$) & (mJy km s$^{-1}$) & (mJy km s$^{-1}$) & (km s$^{-1}$) \\
    (1) & (2) & (3) & (4) & (5) & (6) & (7) \\
    \midrule
    Clump NE & $194 \pm 43$ & $174 \pm 43$ & $148 \pm 36$ & $123 \pm 36$ & $104 \pm 21$ & 61 \\
    Clump SW & $291 \pm 55$ & $303 \pm 66$ & $204 \pm 48$ & $121 \pm 48$ & $122 \pm 23$ & 81 \\
    Clump C (tentative) & $108 \pm 43$ & $113 \pm 29$ & $98 \pm 24$ & $92 \pm 24$ & $93 \pm 17$ & 41 \\
    \bottomrule
    \end{tabular}
    \label{app:fluxes}
    \vspace*{0.2cm}    
    \begin{minipage}{18cm}
    \textbf{Columns}: (1) Clump ID; (2) Flux estimated with method 1 described in Appendix \ref{app:fluxes}; (3) Flux estimated with method 2; (4) Flux estimated with method 3; (5) Flux estimated with method 4; (6) Flux estimated with method 5; (7) Velocity width of the channels encompassing the line emission. \\
    Throughout the paper we adopted fluxes estimated using method 3. All methods are described in Appendix \ref{app:fluxes}.
    \end{minipage}
\end{table*}

\end{appendix}

\end{document}